\documentclass[letterpaper]{article} 

\usepackage{aaai2026}   
\nocopyright            

\usepackage{times}            
\usepackage{helvet}           
\usepackage{courier}          
\usepackage[hyphens]{url}     
\usepackage{graphicx}         
\usepackage{natbib}           
\usepackage{caption}          
\usepackage{subcaption}       
\usepackage{amsmath,amssymb,amsfonts}
\usepackage{bm}
\usepackage{booktabs}
\usepackage{array}
\usepackage{multirow}
\usepackage{algorithm}
\usepackage{algorithmic}
\usepackage{newfloat}
\usepackage{xcolor}

\usepackage{xspace}

\newcommand{\TACO}{\textbf{TACO}\xspace}         
\newcommand{\DAPR}{\textsc{DAPR}\xspace}         
\newcommand{\OGAR}{\textsc{OGAR}\xspace}         
\newcommand{\Ours}{Ours\xspace}                 

\newcommand{\ORM}{\ensuremath{r_{\mathrm{out}}}} 
\newcommand{\Dtool}{\ensuremath{\Delta}}         
\newcommand{\aone}{a_1}                          
\newcommand{\atwo}{a_2}                          
\newcommand{\Afin}{a_f}                          
\newcommand{\imgk}{\mathrm{IMG}}                 
\newcommand{\ystar}{y^{*}}

\newcommand{\Aone}{A_1}                          
\newcommand{\Atwo}{A_2}                          
\newcommand{\Racc}{R_{\mathrm{acc}}}

\newcommand{\Rfmt}{R_{\mathrm{fmt}}}

\newcommand{\eg}{e.g.,\xspace}

\shortcites{liu2026better}

\title{TACO: Tool-Augmented Credit Optimization for Agentic Tool Use}

\author{
    Mingkuan Feng\textsuperscript{\rm 1,}\thanks{Equal Contribution},
    Jinyang Wu\textsuperscript{\rm 1,}\footnotemark[1]\textsuperscript{,}\thanks{Project Leader},
    Hao Gu\textsuperscript{\rm 2},
    Fangrui Lv\textsuperscript{\rm 1},
    Ruihan Jin\textsuperscript{\rm 1}\\
    Chuyuan Zhang\textsuperscript{\rm 1},
    Zhengqi Wen\textsuperscript{\rm 1},
    Jianhua Tao\textsuperscript{\rm 1}
}

\affiliations{
    \textsuperscript{\rm 1}Department of Automation, BNRist, Tsinghua University\\
    \textsuperscript{\rm 2}Institute of Automation, Chinese Academy of Sciences\\
    Corresponding to: fmk24@mails.tsinghua.edu.cn, wu-jy23@mails.tsinghua.edu.cn
}

\begin{document}
\maketitle

\begin{abstract}
Agentic multimodal models perform diverse operations on an image via code and
reason over the returned view, an effective paradigm for fine-grained visual question answering. However, code operations can be useful, redundant, or misleading.
Outcome-only rewards cannot precisely distinguish these cases, and existing process rewards either fail to attribute final correctness to individual tool calls, or require an external judge model. To address this, we introduce
\emph{Tool-Augmented Credit Optimization} (\TACO{}), a \textbf{GRPO} variant for code-tool
agents built on two coupled advantage channels. The first, \emph{Differential Answer-Probe Reward} (\textbf{\DAPR{}}), is a
self-supervised, judge-free tool-contribution advantage that credits each tool call
by its own effect on answering correctly. Probe tokens inserted into the model's reasoning
elicit its predictions with and without the tool, and the difference in outcome
reward is taken as the call's value: positive for a useful call, negative for a
misleading one, and zero for one that changes nothing. This reuses the existing
answer checker with no auxiliary judge, and, being a difference rather than an
absolute probe score, is naturally robust to probe-hacking. The second is the
outcome advantage from the final answer, distributed by \emph{Outcome-Gated
Advantage Routing} (\textbf{\OGAR{}}): a parameter-free rule that, conditioned on the
call's outcome, delivers this credit only to the responsible segments, suppressing
wasted tool calls without any cost term. We train \TACO{} through a two-stage SFT+RL pipeline. Extensive
experiments across perception, reasoning, and general multimodal benchmarks show
that it yields consistent accuracy gains and learns to invoke its tools only when
they help.

\end{abstract}

\section{Introduction}
\label{sec:intro}

\begin{figure}[t]
  \centering
  \includegraphics[width=\columnwidth]{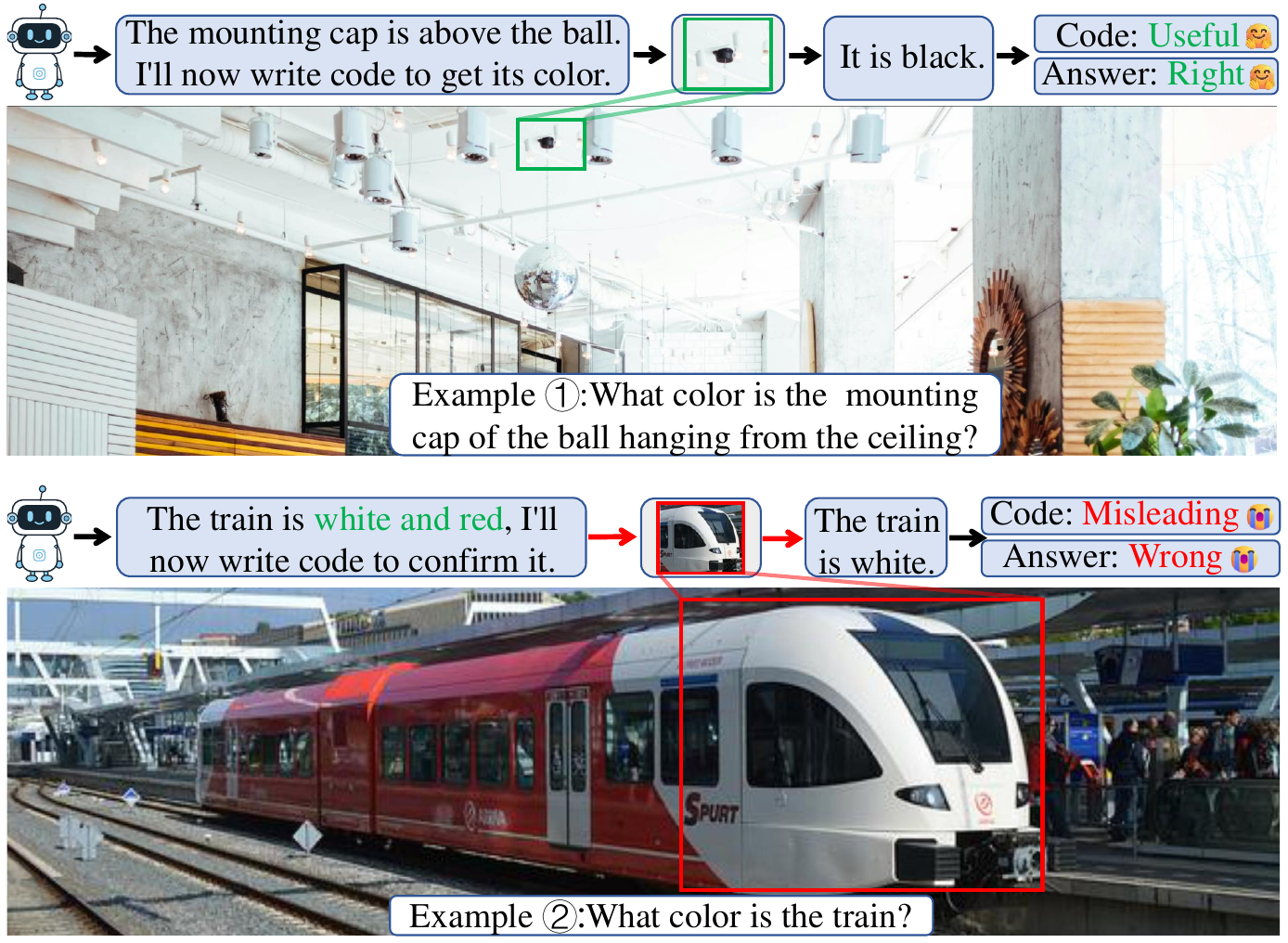}
  \caption{A visual tool call can help or hurt. \textbf{Example~1:} a crop turns a
  wrong answer right (\emph{useful}); \textbf{Example~2:} a crop flips a
  would-be-correct answer to wrong (\emph{misleading}).}
  \label{fig:teaser}
\end{figure}

Recent vision--language models go beyond text-only reasoning to ``think with
images,'' an ability popularized by OpenAI o3~\citep{openai_o3}. They write and execute code that
crops or zooms an image, runs computations, or otherwise transforms the input,
then reason over the
result~\citep{thyme,deepeyes,deepeyesv2,pixelreasoner,minio3,pyvision,mathcodervl,rtwi,hdpo,liu2026better}. When the decisive detail is too small to read,
such a code-tool agent can \emph{act} to obtain a sharper observation rather than
guess~\citep{qi2026patchcue}. These agents are trained with reinforcement learning from verifiable
rewards (RLVR), the standard recipe for eliciting reasoning in LLMs and
VLMs~\citep{deepseekr1,kimik15,kimik25,templaterl}.

However, code operations do not always help~\citep{med,codev}. The same crop can turn a wrong answer right,
leave it unchanged, or turn a right answer wrong (Figure~\ref{fig:teaser}). A tool call is therefore \emph{useful},
\emph{inconclusive}, or \emph{misleading}. To teach an agent to do a code operation when it helps, we need a reward that scores \emph{each call} by its
own contribution, positive for a useful one and \emph{negative} for a harmful one,
and delivers that signal to the tokens that issued the call~\citep{turnlevelreward,toolpo}. The difficulty is that most RLVR optimizes the \emph{final answer} alone, so its
reward attaches to the whole trajectory rather than to the call, and may struggle
to separate a helpful operation from a wasted or a harmful one~\citep{pacr,siop}. Worse, the signal is confounded: a recent analysis
finds that the accuracy gains of crop-zoom tool-use RL are driven mostly by the
model's \emph{intrinsic} improvement rather than by the tool itself~\citep{med},
so a higher final score does not certify that the call helped. Existing process
rewards fall short for two different reasons. Step-wise rewards for
single-chain text reasoning cannot isolate a tool call's
contribution or flag a harmful one~\citep{mig,pacr,spae}; while those defined on a tool's \emph{output}
need an external judge model and never ask whether the call changed the
answer~\citep{codev}. This raises the question we study: \emph{can a self-supervised, judge-free signal
score the tool call by its own effect on final correctness and deliver that
credit only to the tokens responsible for it?}

\begin{figure*}[t]
  \centering
  \includegraphics[width=\textwidth]{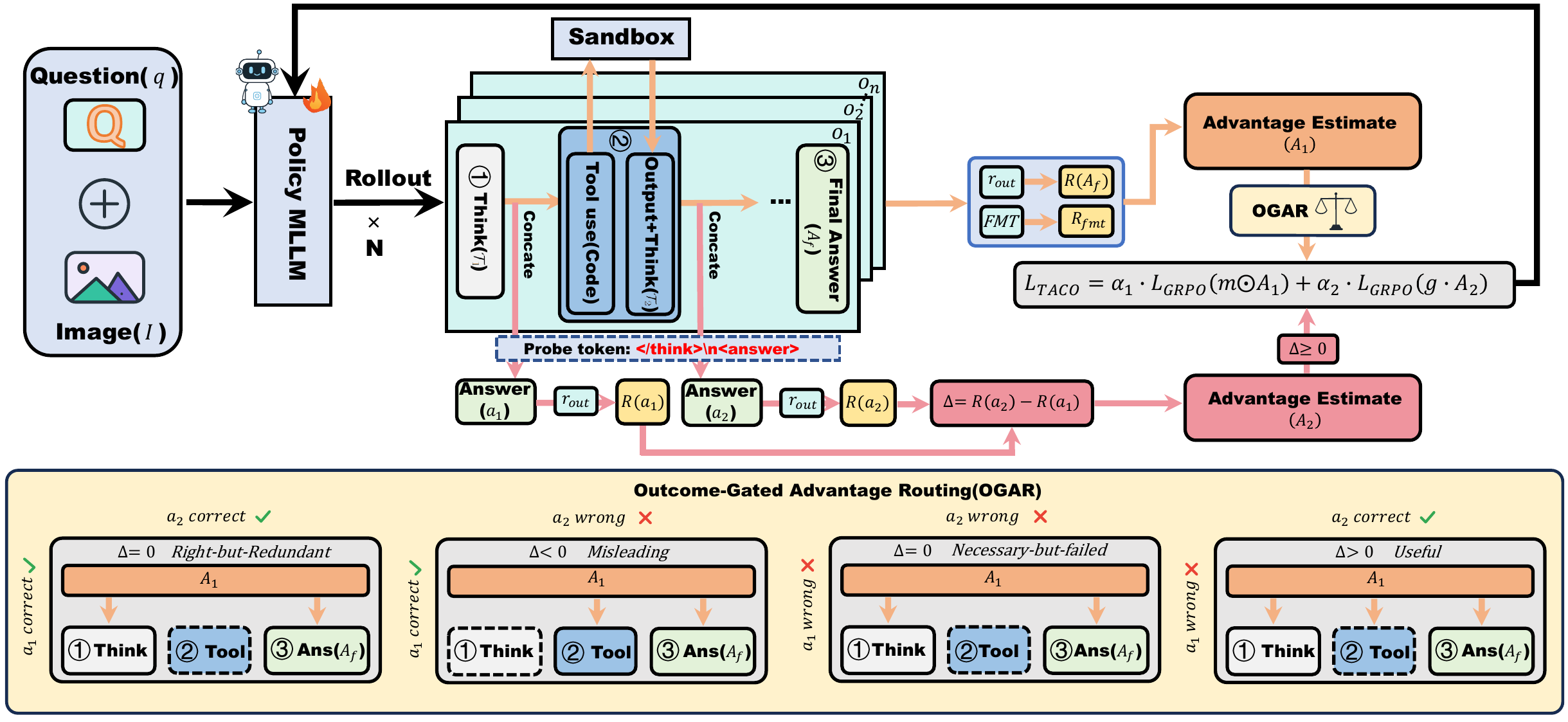}
  \caption{Overview of \TACO{}{}. \textbf{(a)}~The accuracy channel $A_1$ scores the
  final answer; the process channel $A_2$ is the before/after probe difference (the
  tool-call value), gated into the loss only when $\Dtool{\geq}0$. \textbf{(b)}~\OGAR{}
  sends $A_1$ to a segment only when it is responsible for the answer (solid box),
  gating it out otherwise (dashed box).}
  \label{fig:method}
\end{figure*}

We answer this with \emph{Tool-Augmented Credit Optimization} (\TACO{}), a \textbf{GRPO} variant
for code-tool visual agents built on two ingredients (Figure~\ref{fig:method}):
(i)~a tool-call value reward, Differential Answer-Probe Reward
(\textbf{\DAPR{}}), and (ii)~Outcome-Gated Advantage Routing (\textbf{\OGAR{}}) of
the final-answer advantage.

Our key observation is that a tool call splits the trajectory into a clean
\emph{before} and \emph{after}: the increment between them is the entire tool
branch---the code, its observation, and the post-tool reasoning it triggers. We
insert two lightweight probes (Figure~\ref{fig:method}) that read out the agent's
answer just \emph{before} the tool call (tool-off, $\aone$) and the answer it commits
to \emph{after} the tool call has returned and been reasoned over (tool-on, $\atwo$)~\citep{thyme}.
Scoring both with a rule-based answer checker, the \textbf{DAPR} of the call is
their difference: positive for a useful call, negative for a misleading one, and
zero when the call changes nothing. Because the two answers share the same question,
image, and pre-tool reasoning, this difference cancels what the model ``already
knew'' before invoking the tool and credits the call by how much taking the tool
branch changes the answer. \textbf{DAPR} reuses the existing answer checker with no auxiliary judge, and tends to
resist probe-hacking.

A scalar tool-value is not enough: it must reach the \emph{right} tokens. A naive
implementation lets the final-answer advantage land on every token, so a code
block that merely co-occurs with a correct answer is rewarded even when it is redundant, and a correct chain of pre-tool reasoning is blamed for an answer the
tool later spoiled. \textbf{OGAR} fixes this with one principle: \emph{a token segment
receives the final-answer advantage only when it is responsible for the answer,
and responsibility is decided by the call's outcome}. This sorts every
call into four regimes (Figure~\ref{fig:method}): a \emph{useful} call and a
\emph{misleading} one are credited or penalized on the tool branch (the code and
$\mathcal{T}_2$); a \emph{right-but-redundant} call (already correct) has its undue
credit withheld so the wasted call is suppressed; and a \emph{necessary-but-failed}
call (still wrong) has its blame withheld so a warranted attempt on a hard item is
not discouraged. The
gate is parameter-free and needs no tool-call cost term.

We instantiate \TACO{} with a two-stage SFT-then-RL recipe and evaluate across
perception, reasoning, and general multimodal benchmarks, where it delivers
consistent accuracy gains while learning to invoke its tools only when they help. We summarize our main contributions as follows:
\begin{itemize}
  \item \textbf{\TACO{}.} We introduce \TACO{}, a \textbf{GRPO} variant for code-tool visual
  agents that couples \DAPR{} and \OGAR{} into a single objective.
  \item \textbf{Differential Answer-Probe Reward (\DAPR{}).} A self-supervised,
  judge-free, tool-call reward that scores a call by a tool-off/tool-on
  comparison and assigns negative value to misleading calls, unlike per-text-step
  gains, at zero API cost.
  \item \textbf{Outcome-Gated Advantage Routing (\OGAR{}).} A parameter-free,
  token-level rule that routes the final-answer advantage by the call's outcome,
  suppressing wasted calls without any tool-call cost term.
  \item \textbf{Probe-hacking: diagnosis and defense.} We expose a failure mode of
  generative probes, show that the before/after difference is more resistant to it,
  and verify this in the training reward dynamics.
\end{itemize}

\section{Related Work}
\label{sec:related}

\paragraph{Thinking with images.}
Recent work lets multimodal models ``think with images,'' emitting code that crops,
zooms, or transforms the input and reasoning over the returned view.
DeepEyes~\citep{deepeyes}, Pixel-Reasoner~\citep{pixelreasoner}, and
Mini-o3~\citep{minio3} incentivize pixel-space operations with RL;
PyVision~\citep{pyvision}, Thyme~\citep{thyme}, and DeepEyesV2~\citep{deepeyesv2}
run general image-processing code in a sandbox; MathCoder-VL~\citep{mathcodervl}
extends this to math; and Agent0-VL~\citep{agent0vl} pushes toward self-evolving
tool use. Trained almost entirely from \emph{outcome} rewards that credit a
whole trajectory rather than an individual call, these agents tend to over-call their
tools~\citep{hdpo}. MED~\citep{med} shows the apparent gain in crop-and-zoom RL is
largely confounded by the model's own improvement, while
Zoom-Consistency~\citep{zoomconsistency} and RTWI~\citep{rtwi} read intermediate signals
as test-time reliability cues. These observations expose what an outcome reward leaves
implicit: whether a given call actually helped. \textbf{TACO} measures this directly,
scoring each call by its own effect on the answer as a \emph{training} reward rather than
a test-time signal. A code operation splits the trajectory into a clean before and
after that \textbf{TACO} exploits to credit the tool branch itself.

\paragraph{Process rewards and credit assignment.}
A parallel line densifies RL with process rewards on intermediate steps. For
single-chain text reasoning, MIG~\citep{mig} uses a watermarked per-step marginal gain,
PACR~\citep{pacr} rewards progressively ascending confidence, and SPAE~\citep{spae}
estimates step advantages from intermediate confidence and correctness. Defined on a
textual chain in log-probability space, none isolates an external tool observation or
separates a helpful step from a harmful one. For multi-turn LLM agents, SIOP~\citep{siop}
gives a verifier-free turn-level potential (our soundness anchor) and HISR~\citep{hisr}
modulates segmental rewards with hindsight, but neither targets a single visual call;
CodeV~\citep{codev} scores visual tool use but via an external GPT-4o judge, adding API
cost and inheriting its biases. In contrast, \textbf{TACO} scores credit from the agent's own outcome reward on two probe answers: the pre-tool probe gives a
``what would you answer without the tool'' baseline these methods lack, and its signed
before/after difference can penalize misleading calls a marginal-gain signal cannot. The
unit of credit is thus an action with a real observation, not a token span.

\section{Method}
\label{sec:method}

\TACO{} augments Group Relative Policy Optimization (\textbf{GRPO})~\citep{deepseekmath}
for code-tool visual agents with two coupled components (Figure~\ref{fig:method}).
\emph{Differential Answer-Probe Reward} (\DAPR{}, Sec.~\ref{sec:ctv}) scores a tool
call by the change in outcome reward just before versus just after it. \emph{Outcome-Gated
Advantage Routing} (\OGAR{}, Sec.~\ref{sec:dual}) then routes the final-answer advantage
to the responsible tokens through an outcome-conditioned gate. We train in two stages: an
SFT cold-start, then \textbf{GRPO} with this gated dual-channel advantage
(Sec.~\ref{sec:train}).

\subsection{Setting and Notation}
A code-tool visual agent receives a question $q$ and image $I$ and produces a
trajectory that interleaves reasoning with a tool call: it first reasons, then
emits code that is executed in a sandbox and returns a visual observation $\imgk$
(a crop or zoom of $I$), and finally reasons over the result and answers
(Figure~\ref{fig:method}). The trajectory is delimited by three special tokens:
\verb|<think>| for free-form reasoning, \verb|<code>| for a Python program executed
on $I$, and \verb|<answer>| for the final answer that ends the trajectory. We write
$\ORM(\cdot)\in\{-1,0,+1\}$ for the verifiable
outcome reward, a rule-based answer checker (\eg string matching against the
ground truth $\ystar$): $+1$ for a correct answer, $-1$ for an
incorrect one, and $0$ when no answer is produced. We split each
trajectory into three token segments used
throughout: the pre-tool reasoning $\mathcal{T}_1$ (which produces $\aone$), the
code tokens $\mathcal{C}$ of the call, and the post-tool reasoning with the final
answer $\mathcal{T}_2$ (which produces $\Afin$).

\subsection{Differential Answer-Probe Reward (\DAPR{})}
\label{sec:ctv}
Around the tool call we insert two lightweight probes that prefill the answer
header \verb|</think>|\allowbreak\verb|\n<answer>| and greedily decode a short
answer (Figure~\ref{fig:method}):
\begin{itemize}
  \item \textbf{Pre-tool probe (tool-off):} taken after $\texttt{Think}_1$ but
  \emph{before} the code runs; with context $(q,I,\texttt{Think}_1)$, it yields
  the answer $\aone$ the agent would give without invoking the tool.
  \item \textbf{Post-tool probe (tool-on):} taken after the tool call has returned
  and been reasoned over in $\texttt{Think}_2$; the context additionally contains the
  tool view $\imgk$, yielding $\atwo$. With multiple calls this probe is read after
  the final call, so $\atwo$ is the answer the agent commits to once its tool branch
  is complete.
\end{itemize}
Scoring both with $\ORM$ gives the tool-value
\begin{align}
\ORM(\aone) &\in \{-1,0,+1\}, \quad \ORM(\atwo)\in \{-1,0,+1\}, \\
\Dtool &= \ORM(\atwo) - \ORM(\aone).
\label{eq:ctv}
\end{align}
$\Dtool>0$ marks a \emph{useful} call (Figure~\ref{fig:teaser}, Example~1),
$\Dtool<0$ a \emph{misleading} one (Example~2), and $\Dtool=0$ a call that does
not change the outcome: either an \emph{easy} item already correct without the
tool or a \emph{hard} item wrong regardless (\OGAR{} handles these two cases in
Sec.~\ref{sec:dual}). When the
agent emits no code and answers directly, the process channel is empty and only the
accuracy channel applies. Computing $\Dtool$ needs only two short probe decodes and
no API call, unlike CodeV's per-step GPT-4o judge~\citep{codev}.

\paragraph{Differencing cancels the pre-tool baseline.}
$\aone$ and $\atwo$ share the same question, image, and pre-tool reasoning
$\texttt{Think}_1$; the increment between them is the entire tool branch---the code
$\mathcal{C}$, its observation $\imgk$, and the post-tool reasoning
$\texttt{Think}_2$ it triggers. Subtracting therefore cancels what the model
``already knew'' before invoking the tool (including any answer it pre-committed to
in $\texttt{Think}_1$) and credits the call by how much taking the tool branch
changes the answer. By removing this pre-tool baseline, the very term an outcome-only reward leaves in,
$\Dtool$ directly targets the tool-gain confound noted by MED~\citep{med}.

\paragraph{Robustness to probe-hacking.}
\label{sec:hack}
The same cancellation defends against a failure mode of generative probes: a probe
that truncates $\texttt{Think}$ and appends \verb|<answer>| can be exploited if the
model writes its conclusion early into $\texttt{Think}$, since the probe simply
copies it. But both probes read out from the same $\texttt{Think}_1$, so pre-writing
inflates $\ORM(\aone)$ as much as $\ORM(\atwo)$ and the two gains cancel in
$\Dtool$: it lifts the agent's own baseline but leaves $\Dtool$ unchanged. A call
earns positive $\Dtool$ only when taking the tool branch turns a wrong pre-tool
answer right.

\subsection{Outcome-Gated Advantage Routing (\OGAR{})}
\label{sec:dual}
\OGAR{} routes the final-answer advantage to the \emph{right} tokens, using the
tool-value $\Dtool$ as a gate and optimized jointly with the process channel
(Figure~\ref{fig:method}), under one principle: each segment is credited only by the
outcome it controls.

\paragraph{Accuracy channel.}
The final answer $\Afin$ is scored by the rule-based answer checker
$\ORM(\Afin)\in\{-1,0,+1\}$, plus a format term:
\begin{equation}
\Racc = \ORM(\Afin) + 0.5\,\Rfmt .
\end{equation}
Here $\Rfmt$ rewards output that follows the required
\texttt{<think>}/\texttt{<code>}/\texttt{<answer>} structure. Over the group of $G$
rollouts, the accuracy advantage is the standard
\textbf{GRPO} normalization of $\Racc$:
\begin{equation}
\Aone^{(i)} = \frac{\Racc^{(i)} - \mathrm{mean}\big(\{\Racc^{(j)}\}_{j=1}^{G}\big)}
{\mathrm{std}\big(\{\Racc^{(j)}\}_{j=1}^{G}\big)} .
\end{equation}

\paragraph{Process channel.}
The tool-value $\Dtool$ yields a single trajectory-level advantage, normalized over
the group in the same \textbf{GRPO} fashion as the accuracy channel:
\begin{equation}
\Atwo^{(i)} = \frac{\Dtool^{(i)} - \mathrm{mean}\big(\{\Dtool^{(j)}\}_{j=1}^{G}\big)}
{\mathrm{std}\big(\{\Dtool^{(j)}\}_{j=1}^{G}\big)} .
\end{equation}
The process channel acts on the whole sequence, but only for non-misleading calls:
for a misleading call ($\Dtool{<}0$) we switch it off through a trajectory-level gate
$g^{(i)}$ that equals $1$ when $\Dtool^{(i)}{\geq}0$ and $0$ otherwise. The penalty is
then carried entirely by the gated accuracy channel on the code and $\mathcal{T}_2$
(outcome gate below), which keeps $\Aone$ off $\mathcal{T}_1$ when $\Dtool{<}0$, so the
correct pre-tool reasoning $\mathcal{T}_1$ stays unpenalized by both channels.

\paragraph{Outcome gate.}
The final-answer advantage $\Aone$ reaches a segment only when it is
\emph{responsible} for $\Afin$, which the call's outcome $\Dtool$ decides. This
sorts every call into four regimes (Figure~\ref{fig:method}):
\begin{itemize}
  \item \emph{Right-but-redundant} ($\Dtool{=}0$, $\aone$ already correct): the item
  is answered correctly with or without the call, so $\Aone$ is withheld from the
  tool branch (the code and $\mathcal{T}_2$), and the pre-tool reasoning
  $\mathcal{T}_1$ that already solved the item keeps the credit.
  \item \emph{Misleading} ($\Dtool{<}0$): the call misleads the correct pre-tool
  reasoning $\mathcal{T}_1$, so the $\Aone$ blame falls on the code and
  $\mathcal{T}_2$, not on $\mathcal{T}_1$ (Figure~\ref{fig:teaser}, Example~2).
  \item \emph{Necessary-but-failed} ($\Dtool{=}0$, $\aone$ wrong): the $\Aone$
  blame is withheld from the whole tool branch (the code and $\mathcal{T}_2$),
  encouraging exploration of tool use on hard items.
  \item \emph{Useful} ($\Dtool{>}0$): the call positively aids the reasoning, so the
  whole trajectory ($\mathcal{T}_1$, code, and $\mathcal{T}_2$) receives the $\Aone$
  credit.
\end{itemize}
We treat the code and the post-tool reasoning $\mathcal{T}_2$ as a single tool
branch: whenever the call is responsible for $\Afin$ they are gated together. We
write this per-segment routing of $\Aone$ as the gate
\begin{equation}
m[t] =
\begin{cases}
1 & t\in\mathcal{T}_1 \text{ and } \Dtool\geq0,\\
1 & t\in\mathcal{C}\cup\mathcal{T}_2 \text{ and } \Dtool\neq0,\\
0 & \text{otherwise}.
\end{cases}
\end{equation}
which masks $\Aone$ to its responsible tokens while $\Atwo$ covers the whole sequence
whenever it is active ($\Dtool{\geq}0$). In the $\Dtool{=}0$ case the masked tool-branch
tokens still receive $\Atwo$ and the format reward, so the answer credit shifts from a
non-responsible tool branch to the pre-tool reasoning without dropping the signal that
produces the answer. Because the two channels
act on different tokens and at different scales, they are optimized as \emph{separate}
clipped-\textbf{GRPO} losses (Sec.~\ref{sec:train}) and summed, with $m$ the outcome
gate above and $g$ the per-trajectory process gate ($g{=}1$ when $\Dtool{\geq}0$ and
$0$ otherwise) that switches $\Atwo$ off on misleading calls,
\begin{equation}
\mathcal{L}_{\mathrm{TACO}} = \alpha_1\,\mathcal{L}_{\mathrm{GRPO}}(m\odot\Aone)
+ \alpha_2\,\mathcal{L}_{\mathrm{GRPO}}(g\,\Atwo).
\label{eq:loss}
\end{equation}

\begin{table*}[t]
\centering
\caption{Main results across perception, reasoning, and general multimodal
benchmarks. All entries are accuracy in percent. Within the
code-tool / visual-agent group (including Ours), best per column in \textbf{bold}, second
best \underline{underlined}. $^{\dagger}$Instruct variant.}
\label{tab:main}
\setlength{\tabcolsep}{1.5pt}
\renewcommand{\arraystretch}{1.15}
\footnotesize
\begin{tabular}{l cccc ccccc ccc c}
\toprule
\multirow{2}{*}{Model}
 & \multicolumn{4}{c}{Perception}
 & \multicolumn{5}{c}{Reasoning}
 & \multicolumn{3}{c}{General}
 & \multirow{2}{*}{Avg.} \\
\cmidrule(lr){2-5}\cmidrule(lr){6-10}\cmidrule(lr){11-13}
 & HR-4K & HR-8K & MME-RW & V$^{*}$
 & MVision & MVista & MVerse & WeMath & LVista
 & MM$^{*}$ & ChartQA & BLINK & \\
\midrule
\multicolumn{14}{c}{\textit{Closed-source models}}\\
\cmidrule(lr){1-14}
GPT-4o           & 65.0 &59.6 & 62.8&67.5 & 36.5 & 63.4 & 35.3 & 44.2 & 53.2 & 65.7 &85.7 & 63.3 & 58.5 \\
GPT-5            & 75.3 & 74.1 & 68.6 & 72.5 & 61.5 & 81.4 & 66.7 & 77.1 & 69.3 & 73.0 & 76.8 & 69.5 & 72.2 \\
Gemini-2.5-Pro    & 83.9 & 81.5 & 58.3 & 79.1 & 39.8 & 80.9 & 76.9 & 78.0 & 73.8 & 73.6 & 83.6 &  73.5 & 73.6 \\
\midrule
\multicolumn{14}{c}{\textit{Open-source MLLMs (no visual tool)}}\\
\cmidrule(lr){1-14}
LLaVA-OneVision-7B       & 63.0 &59.8 & 57.4 & 75.4 & 17.6 & 58.6 & 19.3 & 20.9 & 33.3 & 61.7 & 80.0 & 48.2 & 49.6 \\
Qwen2.5-VL-7B$^{\dagger}$          & 68.8 &65.3 & 58.3&76.4 & 27.0 & 68.2 & 35.2 & 34.3 & 39.8 & 64.7 &83.7 &56.4 & 56.5 \\
InternVL3-8B             & 70.0 & 69.3 & 61.3 & 70.2& 26.3 & 70.4 & 29.2& 31.7& 45.6 & 68.5& 85.9 & 55.5 & 57.0 \\
Qwen2.5-VL-32B$^{\dagger}$ & 73.4 & 70.4 & 61.0 & 81.2 & 35.2 &72.2 & 40.0 & 47.1 & 54.4& 69.1 &81.1& 63.6 & 62.4 \\
Qwen3-VL-8B$^{\dagger}$     & 78.9 & 74.6 & 60.3 & 86.4 & 53.9 & 77.2& 62.1 & 36.9 & 54.9 &70.9 & 88.1 & 69.1 & 67.8 \\

\midrule
\multicolumn{14}{c}{\textit{Code-tool / visual-agent models}}\\
\cmidrule(lr){1-14}
MathCoder-VL-8B (ACL'25)   & 73.8 &70.6& 52.9 & 77.5 & 26.1& \underline{73.6} &46.5 & \underline{52.1} & 41.6 & 62.5 & 78.8 & 51.7 & 59.0 \\
Pixel-Reasoner-7B (NIPS'25) & 74.0 & 66.9 & 64.4 & 84.3 & 26.3 & 71.2 & 46.9& 36.8 & 46.4 & 64.7 & 76.2 & 56.6 & 59.6 \\
DeepEyes-7B (ICLR'26)       & 75.1 & 72.6 & 64.1 &85.6 & 26.6 & 70.1 & 47.3 & 38.9 & 47.7& 65.3 & 69.4 & 57.5 & 60.0 \\
Mini-o3-7B (ICLR'26)       &77.5 & 73.3 & \underline{65.5} & 88.2 & 25.7 & 67.2 & 45.4 & 34.9 & 45.7 & 63.8& 85.5 & 55.2 & 60.7 \\
Thyme-7B (ICLR'26)        & 77.0 & 72.0 & 64.8& 82.2 & 27.6 & 70.0& 39.1 & 39.3 & 49.0 & 65.9 & 86.1 & 56.1 & 60.8 \\
DeepEyes-v2-7B (ICLR'26)   & 77.9 & 73.8 & 64.9 & 81.8 & 28.9 & 71.9 & 52.7 & 38.1 & 48.7 & 66.1 & 72.2 & 57.8 & 61.2 \\
CodeV-RL-7B (CVPR'26)       & 76.1 & 71.3& 64.5 & 84.8 & \underline{33.6} & 71.8 & 49.2 & 40.5 & 48.7 & \underline{67.6} & 83.5 & 58.2 & 62.5 \\
PyVision-RL-7B (Arxiv'26)& \underline{78.1} & \underline{74.3} & 59.6 & \underline{88.7}& 28.7 & 71.4 & \underline{55.8} & 47.7 & \underline{49.2} & 65.6 & \textbf{86.8} & \underline{58.7} & \underline{63.7} \\
\midrule
\textbf{Ours-7B} & \textbf{83.8} & \textbf{81.6} & \textbf{66.2} & \textbf{89.6} & \textbf{35.8} & \textbf{76.3} & \textbf{57.2} & \textbf{53.1} & \textbf{55.6} & \textbf{69.3} & \underline{86.7} & \textbf{61.6} & \textbf{68.1} \\
\bottomrule
\end{tabular}
\end{table*}

\subsection{Training}
\label{sec:train}
We optimize \TACO{} in two stages.

\paragraph{Stage 1: SFT cold-start.}
Instruction-tuned VLMs rarely invoke code productively out of the box, and RL from
such a start collapses to text-only reasoning~\citep{codev}. We therefore
supervise-fine-tune on trajectories that interleave reasoning, code, and tool
outputs: this teaches the Think--Code--Answer format and provides the reference
policy $\pi_{\mathrm{ref}}$ for Stage~2.

\paragraph{Stage 2: RL with gated advantages.}
For each $(q,I)$ we sample a group of $G$ on-policy rollouts and form the gated
accuracy advantage $m\odot\Aone$ and the process advantage $\Atwo$
(Sec.~\ref{sec:dual}). The importance ratio is
\begin{equation}
r_t = \frac{\pi_\theta(a_t\mid s_t)}{\pi_{\theta_{\mathrm{old}}}(a_t\mid s_t)},
\end{equation}
and the clipped-\textbf{GRPO} surrogate of a per-token advantage $A$ (with $A[t]$ its value
at token $t$) is
\begin{equation}
\mathcal{L}_{\mathrm{GRPO}}(A) = \mathbb{E}_{\tau,t}\!\left[\min\!\big(r_t\,A[t],\,
\mathrm{clip}(r_t,1{-}\epsilon,1{+}\epsilon)\,A[t]\big)\right].
\end{equation}
\TACO{} maximizes $\mathcal{L}_{\mathrm{TACO}}$ (Eq.~\ref{eq:loss}) minus the usual
KL penalty $\beta\,\mathbb{E}[\mathrm{D}_{\mathrm{KL}}(\pi_\theta\,\|\,\pi_{\mathrm{ref}})]$.

\section{Experiments}
\label{sec:exp}

\begin{table*}[t]
\centering
\caption{Accuracy vs.\ end-to-end inference efficiency (7B). For each benchmark we
report accuracy (\%) and per-question latency in seconds (batch~1 on one A100).
The latency gap reflects how many tool/sandbox rounds a policy emits at inference
(CodeV's GPT-4o judge is training-time only). Best accuracy and lowest latency per
benchmark in \textbf{bold}.}
\label{tab:efficiency}
\setlength{\tabcolsep}{4pt}
\renewcommand{\arraystretch}{1.15}
\small
\begin{tabular}{@{}l cc cc cc cc cc@{}}
\toprule
\multirow{2}{*}{Model}
 & \multicolumn{2}{c}{V$^{*}$}
 & \multicolumn{2}{c}{HR-Bench-4K}
 & \multicolumn{2}{c}{HR-Bench-8K}
 & \multicolumn{2}{c}{MathVision}
 & \multicolumn{2}{c}{MMStar} \\
\cmidrule(lr){2-3}\cmidrule(lr){4-5}\cmidrule(lr){6-7}\cmidrule(lr){8-9}\cmidrule(lr){10-11}
 & Acc$\uparrow$ & \shortstack[l]{Latency(s)$\downarrow$} & Acc$\uparrow$ & \shortstack[l]{Latency(s)$\downarrow$} & Acc$\uparrow$ & \shortstack[l]{Latency(s)$\downarrow$} & Acc$\uparrow$ & \shortstack[l]{Latency(s)$\downarrow$} & Acc$\uparrow$ & \shortstack[l]{Latency(s)$\downarrow$} \\
\midrule
Thyme-7B (ICLR'26)         & 82.2 & 3.0 & 77.0 & 3.7 & 72.0 & 4.0 & 27.6 & 2.8 & 65.9 & 2.5 \\
DeepEyes-7B (ICLR'26)      & 85.6 & 2.8 & 75.1 & 5.0 & 72.6 & 5.7 & 26.6 & 3.7 & 65.3 & 3.4 \\
DeepEyes-v2-7B (ICLR'26)   & 81.8 & 3.1 & 77.9 & 5.8 & 73.8 & 5.9& 28.9 & 3.4 & 66.1 & 3.1 \\
CodeV-RL-7B (CVPR'26)      & 84.8 & 3.4 & 76.1 & 5.3 & 71.3 & 6.1 & 33.6 & 2.9 & 67.6 & 2.5 \\
PyVision-RL-7B & 88.7 & 3.6 & 78.1 & 5.5 & 74.3 & 6.2 & 28.7 & 2.9 & 65.6 & 2.6 \\
\midrule
\textbf{Ours-7B} & \textbf{89.6} & \textbf{2.3} & \textbf{83.8} & \textbf{3.2} & \textbf{81.6} & \textbf{3.5} & \textbf{35.8} & \textbf{2.4} & \textbf{69.3} & \textbf{2.0} \\
\bottomrule
\end{tabular}
\end{table*}

We describe our data and implementation below, then report the main comparison
(Sec.~\ref{sec:perf}), an efficiency comparison (Sec.~\ref{sec:eff}), component
ablations (Sec.~\ref{sec:abl}), generalization across base models (Sec.~\ref{sec:gen}),
and training dynamics with the probe-hacking analysis (Sec.~\ref{sec:dyn}).

\subsection{Experimental Setup}
\paragraph{SFT data curation.}
Our SFT data is built on the Thyme SFT corpus~\citep{thyme}, whose trajectories
interleave reasoning, code, and tool outputs. We re-curate it with three filters.
\textbf{(i)~Execution validity}: we re-run every code block in our sandbox and
discard trajectories with execution errors or tool observations/answers
inconsistent with the actual output, which otherwise teach the model to hallucinate
observations. \textbf{(ii)~Tool necessity}: we drop samples that Qwen2.5-VL-7B~\citep{bai2025qwen25vltechnicalreport}
already solves without tools (pass@$8{=}1$), keeping only trajectories where a tool
call is genuinely needed. \textbf{(iii)~Quality}: Gemini-3-Pro scores each
trajectory for reasoning coherence and tool-use rationale, and low-quality or
blind-tool-use traces are removed.

\paragraph{RL data curation.}
For RL we follow CodeV~\citep{codev}, building on its open-source prompt data and
adopting its data-cleaning recipe. Keeping only questions with verifiable
ground-truth answers, we clean them in two ways. \textbf{(i)~Environmental fidelity}: each prompt is passed through
Gemini-3-Pro to check image quality, question clarity, and image-text consistency,
and prompts with corrupted images or severe ambiguity are removed so the policy does
not fit to noise. \textbf{(ii)~Difficulty calibration}: prompts that our SFT checkpoint already solves
on all $G{=}8$ rollouts are trivially easy and yield zero-variance accuracy rewards
(and thus no \textbf{GRPO} advantage), so we remove them.

\paragraph{Implementation details.}
Following Thyme, we build on Qwen2.5-VL-7B~\citep{bai2025qwen25vltechnicalreport}, with $2$ epochs of SFT followed by $1$ epoch of \textbf{GRPO}. The
prompt template specifies the \texttt{<think>}, \texttt{<code>}, \texttt{<answer>},
and sandbox-output tokens. We set $\alpha_1{=}1.0$ and $\alpha_2{=}0.15$
(Sec.~\ref{sec:dual}), use no KL penalty ($\beta{=}0$), and sample $G{=}8$ rollouts
per prompt at temperature $1.0$, with a total batch size of $128$ and learning rate
$1{\times}10^{-6}$. Training runs on a single node of $8{\times}$80\,GB A100 GPUs.
For evaluation we use VLMEvalKit~\citep{vlmevalkit} with its default protocol across
all benchmarks. More details are provided in the Appendix.

\paragraph{Benchmarks and baselines.}
We evaluate on twelve benchmarks in three groups: \emph{perception}
(HR-Bench-4K/8K~\citep{hr4k}, MME-RealWorld~\citep{zhang2025mme}, V$^{*}$~\citep{vstar}), \emph{reasoning} (MathVision~\citep{MathVision}, MathVista~\citep{lu2024mathvista},
MathVerse~\citep{zhang2024mathverse}, WeMath~\citep{qiao2025we}, LogicVista~\citep{xiao2024logicvista}), and \emph{general} (MMStar~\citep{chen2024we}, ChartQA~\citep{masry2022chartqa}, BLINK~\citep{fu2024blink}),
reporting per-benchmark accuracy and the macro-average. Baselines span closed-source
models (GPT-4o, Gemini-2.5-Pro), open-source MLLMs (Qwen2.5-VL,
Qwen2.5-VL-32B-Instruct~\citep{bai2025qwen25vltechnicalreport}, InternVL3~\citep{zhu2025internvl3}, LLaVA-OneVision~\citep{li2024llava}, Qwen3-VL~\citep{bai2025qwen3}), and 7--8B code-tool
agents: Thyme~\citep{thyme},
DeepEyes~\citep{deepeyes}, DeepEyesV2~\citep{deepeyesv2},
Pixel-Reasoner~\citep{pixelreasoner}, Mini-o3~\citep{minio3},
MathCoder-VL~\citep{mathcodervl}, CodeV~\citep{codev}, and PyVision~\citep{pyvision}.
More details are provided in the Appendix.

\subsection{Performance of \TACO{}}
\label{sec:perf}
Table~\ref{tab:main} reports accuracy across the three groups. \TACO{} (``Ours'')
reaches an average of 68.1. The most telling comparison is with the other
\emph{code-tool / visual-agent models}, which share our recipe of acting on the image
through code: \TACO{} clears the best of them (PyVision, 63.7) by $4.4$ points, and
improves on Thyme-7B (60.8), DeepEyes-7B (60.0), DeepEyes-v2-7B (61.2), and
CodeV-7B-RL (62.5) by $5.6$ to $8.1$ points on average, surpassing CodeV's judge-based
process reward without any external judge. Against \emph{closed-source} models, as a 7B
model it substantially outperforms the proprietary GPT-4o, $68.1$ vs.\ $58.5$ ($+9.6$).

The gains concentrate where a code operation can expose a decisive detail. On all four
\emph{perception} benchmarks \TACO{} leads every code-tool agent, raising HR-Bench-8K to
81.6 and V$^{*}$ to 89.6, and it is strongest on the fine-grained \emph{reasoning} tasks
LogicVista (55.6) and WeMath (53.1). Crucially, unlike simply calling the tool more often,
it obtains these gains through \emph{appropriate} tool use
(Table~\ref{tab:efficiency}). We attribute this to the tool-call credit assignment:
\DAPR{} and \OGAR{} reward a call only when it changes a wrong answer to a right one and
withhold credit otherwise, teaching the policy to crop precisely where it helps rather
than over-call.

\begin{figure*}[t]
\centering
\begin{subfigure}{0.33\textwidth}
  \centering
  \includegraphics[width=\textwidth]{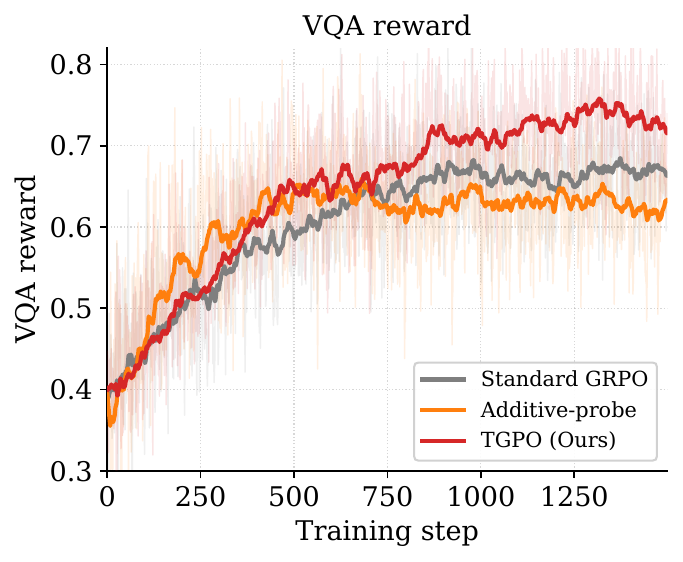}
  \caption{Accuracy reward}
\end{subfigure}\hfill
\begin{subfigure}{0.33\textwidth}
  \centering
  \includegraphics[width=\textwidth]{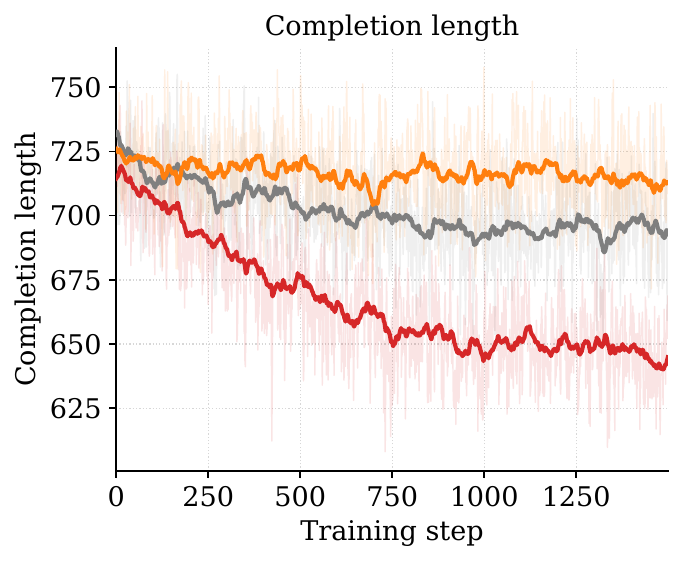}
  \caption{Completion length}
\end{subfigure}\hfill
\begin{subfigure}{0.33\textwidth}
  \centering
  \includegraphics[width=\textwidth]{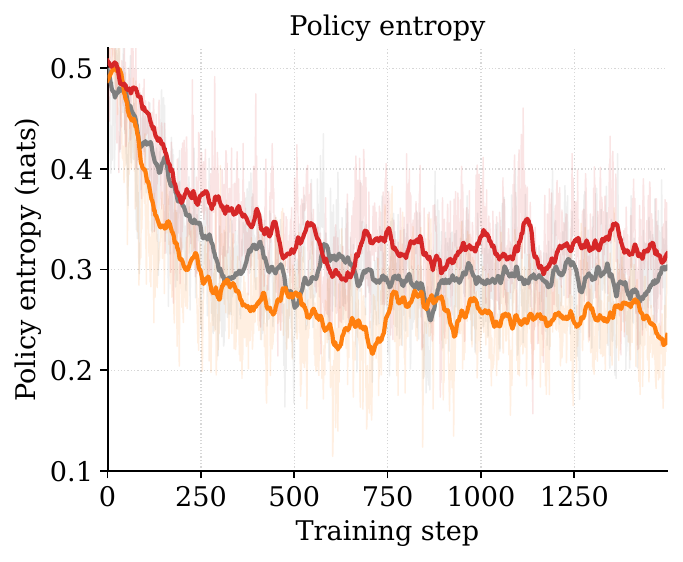}
  \caption{Policy entropy}
\end{subfigure}
\caption{Training dynamics. \textbf{(a)}~Accuracy reward: \TACO{} stays highest, while the
additive-probe variant rises fastest early (probe hacking pays off) but plateaus and ends
below standard \textbf{GRPO}. \textbf{(b)}~Completion length: \TACO{} steadily shortens
completions, ending well below the others, so fewer tool/sandbox rounds mirror its latency
advantage (Table~\ref{tab:efficiency}). \textbf{(c)}~Policy entropy: all three decline
smoothly to a healthy non-zero band without collapsing; \TACO{} keeps the highest entropy
and the additive-probe variant the lowest, matching its premature convergence.}
\label{fig:dynamics}
\end{figure*}

\subsection{Efficiency and Tool Use}
\label{sec:eff}
\TACO{} is simultaneously the most accurate and the fastest agent
(Table~\ref{tab:efficiency}): on all five benchmarks it reaches the highest accuracy
at the lowest end-to-end latency, e.g.\ 89.6\% at 2.3\,s on V$^{*}$ versus 88.7\% at
3.6\,s for PyVision. The speed-up follows directly from the tool-use behavior of
Figure~\ref{fig:dynamics}b: by invoking the tool only when it helps, \TACO{} emits
fewer tool and sandbox rounds per question, so appropriate cropping improves both
accuracy and latency at once. This realizes the goal stated in the abstract, an
agent that uses its tools when they help and abstains when they do not.

\begin{table}[t]
\centering
\caption{Component ablations (accuracy \%, 7B). Best per column in \textbf{bold};
\emph{Avg.} is over the five-benchmark subset shown.
\emph{w/o \DAPR{}} is the additive-probe variant: it keeps the tool-value channel but
rewards the sum $\ORM(\aone){+}\ORM(\atwo)$ instead of the before/after difference.
\emph{w/o \OGAR{}} keeps \DAPR{} but disables gated routing, so the final-answer
advantage $\Aone$ lands on every token rather than only the outcome-selected segments.}
\label{tab:ablation}
\setlength{\tabcolsep}{2pt}
\renewcommand{\arraystretch}{1.2}
\footnotesize
\begin{tabular}{@{}l ccccc c@{}}
\toprule
Method & V$^{*}$ & HR-4K & HR-8K & MathV & MMStar & Avg. \\
\midrule
\textbf{\Ours (full)-7B} & \textbf{89.6} & \textbf{83.8} & \textbf{81.6} & \textbf{35.8} & \textbf{69.3} & \textbf{72.0} \\
\midrule
w/o \DAPR{} & 85.1 & 77.2 & 78.0 & 31.1 & 65.9 & 67.5 \\
w/o \OGAR{} & 87.4 & 81.6 & 80.2 & 33.4 & 67.4 & 70.0 \\
\bottomrule
\end{tabular}
\end{table}

\begin{table}[t]
\centering
\caption{\TACO{} generalizes across base models (accuracy \%, subset of
Table~\ref{tab:main} benchmarks); \emph{Avg.} is over this five-benchmark subset.}
\label{tab:basegen}
\setlength{\tabcolsep}{2pt}
\renewcommand{\arraystretch}{1.2}
\footnotesize
\begin{tabular}{@{}l ccccc c@{}}
\toprule
Base / Method & V$^{*}$ & HR-4K & HR-8K & MathV & MMStar & Avg. \\
\midrule
Qwen2.5-VL-7B & 76.4 & 68.8 & 65.3 & 27.0 & 64.7 & 60.4 \\
\quad + \TACO{} & \textbf{89.6} & \textbf{83.8} & \textbf{81.6} & \textbf{35.8} & \textbf{69.3} & \textbf{72.0} \\
\midrule
Qwen3-VL-8B & 86.4 & 78.9 & 74.6 & 53.9 & 70.9 & 72.9 \\
\quad + \TACO{} & \textbf{92.9} & \textbf{85.3} & \textbf{84.1} & \textbf{59.6} & \textbf{72.1} & \textbf{78.8} \\
\bottomrule
\end{tabular}
\end{table}

\subsection{Ablation Study}
\label{sec:abl}
Table~\ref{tab:ablation} removes one component at a time; the full \TACO{} objective
reaches 72.0, and both components contribute. \emph{w/o \DAPR{}} drops the average to 67.5
($-4.5$): rewarding $\ORM(\aone)$ directly reopens the probe-hacking that differencing
prevents (analysis below). \emph{w/o \OGAR{}} drops it to 70.0 ($-2.0$): spreading the
final-answer advantage over every token blurs the credit that gated routing keeps on the
responsible segments. The two ingredients are complementary, and each is needed for
the full gain.

\subsection{Generalization across base models}
\label{sec:gen}
\TACO{} is not tied to a single backbone. On a representative five-benchmark subset
(Table~\ref{tab:basegen}), it lifts Qwen2.5-VL-7B from 60.4 to 72.0 ($+11.6$; the main
comparison of Table~\ref{tab:main}) and the much stronger Qwen3-VL-8B from 72.9 to 78.8
($+5.9$). The gains are consistent on every benchmark and match what \TACO{} rewards:
largest on high-resolution perception, where deciding \emph{when} and \emph{where} to crop
matters most (HR-8K $+16.3$/$+9.5$, HR-4K $+15.0$/$+6.4$, V$^{*}$ $+13.2$/$+6.5$), and
smaller but never negative on reasoning and general (MathVision $+8.8$/$+5.7$, MMStar
$+4.6$/$+1.2$). The smaller Qwen3-VL-8B lift is expected, as a stronger base leaves less
perceptual headroom; that \TACO{} still adds nearly six points shows the gains come from
the credit-assignment mechanism and compound with, rather than substitute for, a more
capable backbone.

\subsection{Training Dynamics}
\label{sec:dyn}
Figure~\ref{fig:dynamics} compares \TACO{} against two baselines. \emph{Standard
\textbf{GRPO}} trains on the accuracy reward alone, with no tool-value channel and no
gating. The \emph{additive-probe} variant (\emph{w/o \DAPR{}}) adds a tool-value
channel but rewards the two probe scores additively, $\ORM(\aone){+}\ORM(\atwo)$,
instead of taking their difference.

\paragraph{Probe-hacking is real, and differencing defends against it.}
On the accuracy reward (panel a), the additive-probe variant climbs fastest at first, when
writing the answer early into the reasoning inflates its probe score, but it then
plateaus and is overtaken, ending even below standard \textbf{GRPO}: the early spike
was reward, not capability. \TACO{}'s before/after difference cannot be inflated this
way and sustains the highest reward throughout.

\paragraph{The agent learns to crop only when needed.}
Completion length (panel b) falls steadily under \TACO{} (from about $720$ to $640$),
while the additive-probe variant stays longest and standard \textbf{GRPO}
barely moves. We read this as the policy converging to more economical tool use---it
learns to issue a crop only when it expects to help rather than reflexively, which
shortens the average response. The same converged policy is the one we evaluate, so this
learned economy is consistent with its lower end-to-end latency in
Table~\ref{tab:efficiency}.

\paragraph{Exploration is preserved, not collapsed.}
Policy entropy (panel c) tells the same story from the other side: all three variants
start from a similar level and decline smoothly without collapsing, but the additive-probe
variant falls fastest as it commits to the shortcut, while \TACO{} keeps the highest
entropy. Because
\DAPR{} rewards a genuine change in the answer rather than a gameable probe score, the
policy has no shortcut to collapse onto and keeps exploring, which sustains its late
reward growth in panel a.

\section{Conclusion}
We presented \emph{Tool-Augmented Credit Optimization} (\TACO{}), a \textbf{GRPO} variant
that gives code-tool visual agents a tool-call learning signal. It couples \DAPR{}, a
judge-free, probe-hacking-robust reward that scores each call by the before/after
difference of the model's own answer-probe outcomes, with \OGAR{}, which routes the
final-answer advantage only to the responsible segments, reinforcing useful calls and
suppressing wasted ones. Across twelve benchmarks, \TACO{} achieves
the best average among prior code-tool agents,
and transfers to a stronger Qwen3-VL backbone, showing the gains come from the mechanism,
not the base or data. The bottleneck is learning which tool calls are worth making, which \TACO{} scores at the tool-call level.
More broadly, this value can be read from the agent's \emph{own} outcome reward by
differencing a before/after answer probe, with no auxiliary judge or cost term.

\bibliography{references}

\begin{thebibliography}{53}
\providecommand{\natexlab}[1]{#1}

\bibitem[{Bai et~al.(2025{\natexlab{a}})Bai, Cai, Chen, Chen, Chen, Cheng,
  Deng, Ding, Gao, Ge et~al.}]{bai2025qwen3}
Bai, S.; Cai, Y.; Chen, R.; Chen, K.; Chen, X.; Cheng, Z.; Deng, L.; Ding, W.;
  Gao, C.; Ge, C.; et~al. 2025{\natexlab{a}}.
\newblock Qwen3-vl technical report.
\newblock \emph{arXiv preprint arXiv:2511.21631}.

\bibitem[{Bai et~al.(2025{\natexlab{b}})Bai, Chen, Liu, Wang, Ge, Song, Dang,
  Wang, Wang, Tang, Zhong, Zhu, Yang, Li, Wan, Wang, Ding, Fu, Xu, Ye, Zhang,
  Xie, Cheng, Zhang, Yang, Xu, and Lin}]{bai2025qwen25vltechnicalreport}
Bai, S.; Chen, K.; Liu, X.; Wang, J.; Ge, W.; Song, S.; Dang, K.; Wang, P.;
  Wang, S.; Tang, J.; Zhong, H.; Zhu, Y.; Yang, M.; Li, Z.; Wan, J.; Wang, P.;
  Ding, W.; Fu, Z.; Xu, Y.; Ye, J.; Zhang, X.; Xie, T.; Cheng, Z.; Zhang, H.;
  Yang, Z.; Xu, H.; and Lin, J. 2025{\natexlab{b}}.
\newblock Qwen2.5-VL Technical Report.
\newblock arXiv:2502.13923.

\bibitem[{Chen et~al.(2024)Chen, Li, Dong, Zhang, Zang, Chen, Duan, Wang, Qiao,
  Lin et~al.}]{chen2024we}
Chen, L.; Li, J.; Dong, X.; Zhang, P.; Zang, Y.; Chen, Z.; Duan, H.; Wang, J.;
  Qiao, Y.; Lin, D.; et~al. 2024.
\newblock Are we on the right way for evaluating large vision-language models?
\newblock \emph{Advances in Neural Information Processing Systems}, 37:
  27056--27087.

\bibitem[{Duan et~al.(2024)Duan, Yang, Qiao, Fang, Chen, Liu, Dong, Zang,
  Zhang, Wang, Lin, and Chen}]{vlmevalkit}
Duan, H.; Yang, J.; Qiao, Y.; Fang, X.; Chen, L.; Liu, Y.; Dong, X.; Zang, Y.;
  Zhang, P.; Wang, J.; Lin, D.; and Chen, K. 2024.
\newblock VLMEvalKit: An Open-Source Toolkit for Evaluating Large
  Multi-Modality Models.
\newblock In \emph{Proceedings of the 32nd ACM International Conference on
  Multimedia}, 11198--11201.

\bibitem[{Feng et~al.(2026)Feng, Wu, Liu, Zhang, Fang, Jin, Che, Shao, Wen, and
  Tao}]{feng2026two}
Feng, M.; Wu, J.; Liu, S.; Zhang, S.; Fang, H.; Jin, R.; Che, F.; Shao, P.;
  Wen, Z.; and Tao, J. 2026.
\newblock Two-stage regularization-based structured pruning for llms.
\newblock In \emph{Proceedings of the 64th Annual Meeting of the Association
  for Computational Linguistics (Volume 1: Long Papers)}, 2996--3012.

\bibitem[{Feng et~al.(2025)Feng, Wu, Zhang, Shao, Jin, Wen, Tao, and
  Che}]{feng2025dress}
Feng, M.; Wu, J.; Zhang, S.; Shao, P.; Jin, R.; Wen, Z.; Tao, J.; and Che, F.
  2025.
\newblock Dress: Data-driven regularized structured streamlining for large
  language models.
\newblock \emph{arXiv preprint arXiv:2501.17905}.

\bibitem[{Fu et~al.(2024)Fu, Hu, Li, Feng, Wang, Lin, Roth, Smith, Ma, and
  Krishna}]{fu2024blink}
Fu, X.; Hu, Y.; Li, B.; Feng, Y.; Wang, H.; Lin, X.; Roth, D.; Smith, N.~A.;
  Ma, W.-C.; and Krishna, R. 2024.
\newblock Blink: Multimodal large language models can see but not perceive.
\newblock In \emph{European Conference on Computer Vision}, 148--166. Springer.

\bibitem[{Guo et~al.(2025)Guo, Yang, Zhang, Song, Wang, Zhu, Xu, Zhang, Ma, Bi
  et~al.}]{deepseekr1}
Guo, D.; Yang, D.; Zhang, H.; Song, J.; Wang, P.; Zhu, Q.; Xu, R.; Zhang, R.;
  Ma, S.; Bi, X.; et~al. 2025.
\newblock Deepseek-r1: Incentivizing reasoning capability in llms via
  reinforcement learning.
\newblock \emph{arXiv preprint arXiv:2501.12948}.

\bibitem[{Hong et~al.(2026)Hong, Zhao, Zhu, Lu, Xu, and XingYu}]{deepeyesv2}
Hong, J.; Zhao, C.; Zhu, C.; Lu, W.; Xu, G.; and XingYu. 2026.
\newblock DeepEyesV2: Toward Agentic Multimodal Model.
\newblock In \emph{The Fourteenth International Conference on Learning
  Representations}.

\bibitem[{Hou et~al.(2026)Hou, Xu, Biyani, Li, Liu, Hollon, and Wang}]{codev}
Hou, X.; Xu, S.; Biyani, M.; Li, M.; Liu, J.; Hollon, T.~C.; and Wang, B. 2026.
\newblock Codev: Code with images for faithful visual reasoning via tool-aware
  policy optimization.
\newblock In \emph{Proceedings of the IEEE/CVF Conference on Computer Vision
  and Pattern Recognition}, 21500--21510.

\bibitem[{Hu et~al.(2026)Hu, Dai, Han, Fang, Zhao, Kwong, and Fang}]{siop}
Hu, S.; Dai, Y.; Han, X.; Fang, Z.; Zhao, Y.; Kwong, S. T.~W.; and Fang, Y.
  2026.
\newblock {Self-Induced Outcome Potential: Turn-Level Credit Assignment for
  Agents without Verifiers}.
\newblock \emph{arXiv preprint arXiv:2605.04984}.

\bibitem[{Jin et~al.(2026)Jin, Shao, Wen, Wu, Feng, Yang, Zhang, and
  Tao}]{jin2026exploring}
Jin, R.; Shao, P.; Wen, Z.; Wu, J.; Feng, M.; Yang, S.; Zhang, C.~Y.; and Tao,
  J. 2026.
\newblock Exploring Knowledge Purification in Multi-Teacher Knowledge
  Distillation for LLMs.
\newblock \emph{arXiv preprint arXiv:2602.01064}.

\bibitem[{Jin et~al.(2025)Jin, Shao, Wen, Wu, Feng, Zhang, and
  Tao}]{jin2025radialrouter}
Jin, R.; Shao, P.; Wen, Z.; Wu, J.; Feng, M.; Zhang, S.; and Tao, J. 2025.
\newblock Radialrouter: Structured representation for efficient and robust
  large language models routing.
\newblock \emph{arXiv preprint arXiv:2506.03880}.

\bibitem[{Kim and Chelikavada(2026)}]{zoomconsistency}
Kim, K.; and Chelikavada, K. 2026.
\newblock {Zoom Consistency: A Free Confidence Signal in Multi-Step Visual
  Grounding Pipelines}.
\newblock \emph{arXiv preprint arXiv:2604.15376}.

\bibitem[{Lai et~al.(2026)Lai, Li, Li, Liu, Li, and Zhao}]{minio3}
Lai, X.; Li, J.; Li, W.; Liu, T.; Li, T.; and Zhao, H. 2026.
\newblock Mini-o3: Scaling Up Reasoning Patterns and Interaction Turns for
  Visual Search.
\newblock In \emph{The Fourteenth International Conference on Learning
  Representations}.

\bibitem[{Li et~al.(2024)Li, Zhang, Guo, Zhang, Li, Zhang, Zhang, Zhang, Li,
  Liu et~al.}]{li2024llava}
Li, B.; Zhang, Y.; Guo, D.; Zhang, R.; Li, F.; Zhang, H.; Zhang, K.; Zhang, P.;
  Li, Y.; Liu, Z.; et~al. 2024.
\newblock Llava-onevision: Easy visual task transfer.
\newblock \emph{arXiv preprint arXiv:2408.03326}.

\bibitem[{Li et~al.(2026{\natexlab{a}})Li, Yang, Lin, Dai, Yang, and
  Peng}]{rtwi}
Li, H.; Yang, Y.; Lin, Y.; Dai, X.; Yang, M.; and Peng, X. 2026{\natexlab{a}}.
\newblock Reliable Thinking with Images.
\newblock In \emph{International Conference on Machine Learning}.

\bibitem[{Li et~al.(2026{\natexlab{b}})Li, Jiao, Jin, Dong, Jin, Wang, Wang,
  Zhu, Wen, Lu, and Dou}]{toolpo}
Li, X.; Jiao, W.; Jin, J.; Dong, G.; Jin, J.; Wang, Y.; Wang, H.; Zhu, Y.; Wen,
  J.-R.; Lu, Y.; and Dou, Z. 2026{\natexlab{b}}.
\newblock DeepAgent: A General Reasoning Agent with Scalable Toolsets.
\newblock arXiv:2510.21618.

\bibitem[{Liu, Feng, and Chen(2026)}]{liu2026better}
Liu, J.; Feng, M.; and Chen, L. 2026.
\newblock Better, stronger, faster: Tackling the trilemma in mllm-based
  segmentation with simultaneous textual mask prediction.
\newblock In \emph{Proceedings of the IEEE/CVF Conference on Computer Vision
  and Pattern Recognition}, 33121--33130.

\bibitem[{Liu et~al.(2025)Liu, Xiong, Xia, Zhou, Ji, Feng, Han, Ding, and
  Yao}]{agent0vl}
Liu, J.; Xiong, K.; Xia, P.; Zhou, Y.; Ji, H.; Feng, L.; Han, S.; Ding, M.; and
  Yao, H. 2025.
\newblock {Agent0-VL: Exploring Self-Evolving Agent for Tool-Integrated
  Vision-Language Reasoning}.
\newblock \emph{arXiv preprint arXiv:2511.19900}.

\bibitem[{Lu et~al.(2024)Lu, Bansal, Xia, Liu, Li, Hajishirzi, Cheng, Chang,
  Galley, and Gao}]{lu2024mathvista}
Lu, P.; Bansal, H.; Xia, T.; Liu, J.; Li, C.; Hajishirzi, H.; Cheng, H.; Chang,
  K.-W.; Galley, M.; and Gao, J. 2024.
\newblock Mathvista: Evaluating mathematical reasoning of foundation models in
  visual contexts.
\newblock In \emph{International Conference on Learning Representations},
  volume 2024, 23439--23554.

\bibitem[{Lu et~al.(2026)Lu, Lin, Jia, Tian, Ye, Li, Jin, Liu, Xu, and
  Feng}]{hisr}
Lu, Z.; Lin, Z.; Jia, W.; Tian, C.; Ye, D.; Li, P.; Jin, L.; Liu, N.; Xu, G.;
  and Feng, W. 2026.
\newblock {HISR: Hindsight Information Modulated Segmental Process Rewards for
  Multi-turn Agentic Reinforcement Learning}.
\newblock \emph{arXiv preprint arXiv:2603.18683}.

\bibitem[{Ma et~al.(2026)Ma, Zhang, Li, Du, Shen, and Liu}]{med}
Ma, Y.; Zhang, W.; Li, T.; Du, L.; Shen, X.; and Liu, P. 2026.
\newblock {What Does Vision Tool-Use Reinforcement Learning Really Learn?
  Disentangling Tool-Induced and Intrinsic Effects for Crop-and-Zoom}.
\newblock \emph{arXiv preprint arXiv:2602.01334}.
\newblock ICML 2026.

\bibitem[{Masry et~al.(2022)Masry, Do, Tan, Joty, and Hoque}]{masry2022chartqa}
Masry, A.; Do, X.~L.; Tan, J.~Q.; Joty, S.; and Hoque, E. 2022.
\newblock Chartqa: A benchmark for question answering about charts with visual
  and logical reasoning.
\newblock In \emph{Findings of the association for computational linguistics:
  ACL 2022}, 2263--2279.

\bibitem[{Ng, Harada, and Russell(1999)}]{ng1999policy}
Ng, A.~Y.; Harada, D.; and Russell, S. 1999.
\newblock Policy Invariance Under Reward Transformations: Theory and
  Application to Reward Shaping.
\newblock In \emph{International Conference on Machine Learning (ICML)},
  278--287.

\bibitem[{{OpenAI}(2025)}]{openai_o3}
{OpenAI}. 2025.
\newblock {Thinking with Images}.
\newblock \url{https://openai.com/index/thinking-with-images/}.

\bibitem[{Qi et~al.(2026)Qi, Fu, Li, Liu, Jiang, Qin, Luo, and
  Luan}]{qi2026patchcue}
Qi, Y.; Fu, P.; Li, H.; Liu, Y.; Jiang, C.; Qin, B.; Luo, Z.; and Luan, J.
  2026.
\newblock Patchcue: Enhancing vision-language model reasoning with patch-based
  visual cues.
\newblock \emph{arXiv preprint arXiv:2603.05869}.

\bibitem[{Qiao et~al.(2025)Qiao, Tan, Dong, MinhuiWu, Sun, Song, Wang, Gongque,
  Lei, Zhang et~al.}]{qiao2025we}
Qiao, R.; Tan, Q.; Dong, G.; MinhuiWu, M.; Sun, C.; Song, X.; Wang, J.;
  Gongque, Z.; Lei, S.; Zhang, Y.; et~al. 2025.
\newblock We-math: Does your large multimodal model achieve human-like
  mathematical reasoning?
\newblock In \emph{Proceedings of the 63rd Annual Meeting of the Association
  for Computational Linguistics (Volume 1: Long Papers)}, 20023--20070.

\bibitem[{Shao et~al.(2024)Shao, Wang, Zhu, Xu, Song, Bi, Zhang, Zhang, Li, Wu,
  and Guo}]{deepseekmath}
Shao, Z.; Wang, P.; Zhu, Q.; Xu, R.; Song, J.; Bi, X.; Zhang, H.; Zhang, M.;
  Li, Y.~K.; Wu, Y.; and Guo, D. 2024.
\newblock {DeepSeekMath: Pushing the Limits of Mathematical Reasoning in Open
  Language Models}.
\newblock \emph{arXiv preprint arXiv:2402.03300}.

\bibitem[{Team et~al.(2026)Team, Bai, Bai, Bao, Cai, Cao, Charles, Che, Chen,
  Chen et~al.}]{kimik25}
Team, K.; Bai, T.; Bai, Y.; Bao, Y.; Cai, S.; Cao, Y.; Charles, Y.; Che, H.;
  Chen, C.; Chen, G.; et~al. 2026.
\newblock Kimi K2. 5: Visual Agentic Intelligence.
\newblock \emph{arXiv preprint arXiv:2602.02276}.

\bibitem[{Team et~al.(2025)Team, Du, Gao, Xing, Jiang, Chen, Li, Xiao, Du, Liao
  et~al.}]{kimik15}
Team, K.; Du, A.; Gao, B.; Xing, B.; Jiang, C.; Chen, C.; Li, C.; Xiao, C.; Du,
  C.; Liao, C.; et~al. 2025.
\newblock Kimi k1. 5: Scaling reinforcement learning with llms.
\newblock \emph{arXiv preprint arXiv:2501.12599}.

\bibitem[{Wang et~al.(2025{\natexlab{a}})Wang, Su, Ren, Lin, and
  Chen}]{pixelreasoner}
Wang, H.; Su, A.; Ren, W.; Lin, F.; and Chen, W. 2025{\natexlab{a}}.
\newblock Pixel Reasoner: Incentivizing Pixel-Space Reasoning with
  Curiosity-Driven Reinforcement Learning.
\newblock arXiv:2505.15966.

\bibitem[{Wang et~al.(2024)Wang, Pan, Shi, Lu, Ren, Zhou, Zhan, and
  Li}]{MathVision}
Wang, K.; Pan, J.; Shi, W.; Lu, Z.; Ren, H.; Zhou, A.; Zhan, M.; and Li, H.
  2024.
\newblock Measuring multimodal mathematical reasoning with math-vision dataset.
\newblock \emph{Advances in Neural Information Processing Systems}, 37:
  95095--95169.

\bibitem[{Wang et~al.(2025{\natexlab{b}})Wang, Pan, Wei, Zhou, Shi, Lu, Xiao,
  Yang, Ren, Zhan et~al.}]{mathcodervl}
Wang, K.; Pan, J.; Wei, L.; Zhou, A.; Shi, W.; Lu, Z.; Xiao, H.; Yang, Y.; Ren,
  H.; Zhan, M.; et~al. 2025{\natexlab{b}}.
\newblock Mathcoder-vl: Bridging vision and code for enhanced multimodal
  mathematical reasoning.
\newblock In \emph{Findings of the Association for Computational Linguistics:
  ACL 2025}, 2505--2534.

\bibitem[{Wang et~al.(2025{\natexlab{c}})Wang, Ding, Zeng, Zhou, Shen, Luo, Yu,
  and Tao}]{hr4k}
Wang, W.; Ding, L.; Zeng, M.; Zhou, X.; Shen, L.; Luo, Y.; Yu, W.; and Tao, D.
  2025{\natexlab{c}}.
\newblock Divide, conquer and combine: A training-free framework for
  high-resolution image perception in multimodal large language models.
\newblock In \emph{Proceedings of the AAAI Conference on Artificial
  Intelligence}, volume~39, 7907--7915.

\bibitem[{Wang et~al.(2026)Wang, Wang, Chen, Nimalsiri, and Halgamuge}]{mig}
Wang, X.; Wang, W.; Chen, K.; Nimalsiri, N.; and Halgamuge, S. 2026.
\newblock {Discovering Process-Outcome Credit in Multi-Step LLM Reasoning}.
\newblock \emph{arXiv preprint arXiv:2602.01034}.

\bibitem[{Wei et~al.(2025)Wei, Zeng, Li, Brown, Frunza, Deng, Schneider,
  Nevmyvaka, Zhao, Garcia, and Hong}]{turnlevelreward}
Wei, Q.; Zeng, S.; Li, C.; Brown, W.; Frunza, O.; Deng, W.; Schneider, A.;
  Nevmyvaka, Y.; Zhao, Y.~K.; Garcia, A.; and Hong, M. 2025.
\newblock {Reinforcing Multi-Turn Reasoning in LLM Agents via Turn-Level Reward
  Design}.
\newblock \emph{arXiv preprint arXiv:2505.11821}.

\bibitem[{Wu et~al.(2026{\natexlab{a}})Wu, Zhang, Chang, Zhang, Liu, and
  Du}]{spae}
Wu, F.; Zhang, Z.; Chang, Q.; Zhang, J.; Liu, Q.; and Du, J.
  2026{\natexlab{a}}.
\newblock {Step Potential Advantage Estimation: Harnessing Intermediate
  Confidence and Correctness for Efficient Mathematical Reasoning}.
\newblock \emph{arXiv preprint arXiv:2601.03823}.

\bibitem[{Wu et~al.(2026{\natexlab{b}})Wu, Feng, Zhai, Zhang, Lian, Lv, Shao,
  Jin, Wen, and Tao}]{wu2026astar}
Wu, J.; Feng, M.; Zhai, G.; Zhang, S.; Lian, Z.; Lv, F.; Shao, P.; Jin, R.;
  Wen, Z.; and Tao, J. 2026{\natexlab{b}}.
\newblock Astar: Boosting multimodal reasoning with automated structured
  thinking.
\newblock In \emph{Proceedings of the AAAI Conference on Artificial
  Intelligence}, volume~40, 33926--33934.

\bibitem[{Wu et~al.(2026{\natexlab{c}})Wu, Feng, Zhang, Che, Wen, Liao, Yang,
  Luo, Lian, and Tao}]{wu2026beyond}
Wu, J.; Feng, M.; Zhang, S.; Che, F.; Wen, Z.; Liao, C.; Yang, L.; Luo, H.;
  Lian, Z.; and Tao, J. 2026{\natexlab{c}}.
\newblock Beyond Examples: Towards Automated Thought-level In-Context Reasoning
  for Large Language Models.
\newblock In \emph{Proceedings of the 64th Annual Meeting of the Association
  for Computational Linguistics (Volume 1: Long Papers)}, 2955--2995.

\bibitem[{Wu et~al.(2026{\natexlab{d}})Wu, Liao, Feng, Zhang, Wen, Luo, Yang,
  Xu, and Tao}]{templaterl}
Wu, J.; Liao, C.; Feng, M.; Zhang, S.; Wen, Z.; Luo, H.; Yang, L.; Xu, H.; and
  Tao, J. 2026{\natexlab{d}}.
\newblock {TemplateRL: Structured Template-Guided Reinforcement Learning for
  LLM Reasoning}.
\newblock \emph{arXiv preprint arXiv:2505.15692}.

\bibitem[{Wu et~al.(2025{\natexlab{a}})Wu, Liao, Feng, Zhang, Wen, Shao, Xu,
  and Tao}]{wu2025thought}
Wu, J.; Liao, C.; Feng, M.; Zhang, S.; Wen, Z.; Shao, P.; Xu, H.; and Tao, J.
  2025{\natexlab{a}}.
\newblock Thought-augmented policy optimization: Bridging external guidance and
  internal capabilities.
\newblock \emph{arXiv preprint arXiv:2505.15692}, 1(8): 10.

\bibitem[{Wu et~al.(2025{\natexlab{b}})Wu, Zhang, Che, Feng, Shao, and
  Tao}]{wu2025pandora}
Wu, J.; Zhang, S.; Che, F.; Feng, M.; Shao, P.; and Tao, J. 2025{\natexlab{b}}.
\newblock Pandora's box or aladdin's lamp: A comprehensive analysis revealing
  the role of rag noise in large language models.
\newblock In \emph{Proceedings of the 63rd Annual Meeting of the Association
  for Computational Linguistics (Volume 1: Long Papers)}, 5019--5039.

\bibitem[{Wu and Xie(2024)}]{vstar}
Wu, P.; and Xie, S. 2024.
\newblock V?: Guided visual search as a core mechanism in multimodal llms.
\newblock In \emph{Proceedings of the IEEE/CVF Conference on Computer Vision
  and Pattern Recognition}, 13084--13094.

\bibitem[{Xiao et~al.(2024)Xiao, Sun, Liu, and Wang}]{xiao2024logicvista}
Xiao, Y.; Sun, E.; Liu, T.; and Wang, W. 2024.
\newblock Logicvista: Multimodal llm logical reasoning benchmark in visual
  contexts.
\newblock \emph{arXiv preprint arXiv:2407.04973}.

\bibitem[{Yan et~al.(2026)Yan, Tong, Xue, Tang, Wang, Shi, Zhang, Li, and
  Zou}]{hdpo}
Yan, S.; Tong, J.; Xue, H.; Tang, X.; Wang, Y.; Shi, K.; Zhang, G.; Li, R.; and
  Zou, Y. 2026.
\newblock {Act Wisely: Cultivating Meta-Cognitive Tool Use in Agentic
  Multimodal Models}.
\newblock \emph{arXiv preprint arXiv:2604.08545}.

\bibitem[{Yoon et~al.(2025)Yoon, Yoon, Jang, Eom, Dai, Luo, Hasegawa-Johnson,
  and Yoo}]{pacr}
Yoon, E.; Yoon, H.~S.; Jang, J.; Eom, S.; Dai, Q.; Luo, C.; Hasegawa-Johnson,
  M.~A.; and Yoo, C.~D. 2025.
\newblock {PACR: Progressively Ascending Confidence Reward for LLM Reasoning}.
\newblock \emph{arXiv preprint arXiv:2510.22255}.

\bibitem[{Zhang et~al.(2024)Zhang, Jiang, Zhang, Lin, Guo, Qiu, Zhou, Lu,
  Chang, Qiao et~al.}]{zhang2024mathverse}
Zhang, R.; Jiang, D.; Zhang, Y.; Lin, H.; Guo, Z.; Qiu, P.; Zhou, A.; Lu, P.;
  Chang, K.-W.; Qiao, Y.; et~al. 2024.
\newblock Mathverse: Does your multi-modal llm truly see the diagrams in visual
  math problems?
\newblock In \emph{European Conference on Computer Vision}, 169--186. Springer.

\bibitem[{Zhang et~al.(2026)Zhang, Lu, Yin, Fu, Chen, Hu, Wen, Jiang, Liu,
  Zhang, fan, Chen, Chen, Ding, Tang, Zhang, Wang, Yang, Gao, and Zhou}]{thyme}
Zhang, Y.; Lu, X.; Yin, S.; Fu, C.; Chen, W.; Hu, X.; Wen, B.; Jiang, K.; Liu,
  C.; Zhang, T.; fan, H.; Chen, K.; Chen, J.; Ding, H.; Tang, K.; Zhang, Z.;
  Wang, L.; Yang, F.; Gao, T.; and Zhou, G. 2026.
\newblock Thyme: Think Beyond Images.
\newblock In \emph{The Fourteenth International Conference on Learning
  Representations}.

\bibitem[{Zhang et~al.(2025)Zhang, Zhang, Tian, Fu, Zhang, Wu, Li, Wang, Wen,
  Zhang et~al.}]{zhang2025mme}
Zhang, Y.; Zhang, H.; Tian, H.; Fu, C.; Zhang, S.; Wu, J.; Li, F.; Wang, K.;
  Wen, Q.; Zhang, Z.; et~al. 2025.
\newblock Mme-realworld: Could your multimodal llm challenge high-resolution
  real-world scenarios that are difficult for humans?
\newblock In \emph{International Conference on Learning Representations},
  volume 2025, 89655--89701.

\bibitem[{Zhao et~al.(2026)Zhao, Lin, Li, Zhang, Peng, Zhang, and
  Wei}]{pyvision}
Zhao, S.; Lin, S.; Li, M.; Zhang, H.; Peng, W.; Zhang, K.; and Wei, C. 2026.
\newblock PyVision-RL: Forging Open Agentic Vision Models via RL.
\newblock arXiv:2602.20739.

\bibitem[{Zheng et~al.(2026)Zheng, Yang, Hong, Zhao, Xu, Yang, Shen, and
  XingYu}]{deepeyes}
Zheng, Z.; Yang, M.; Hong, J.; Zhao, C.; Xu, G.; Yang, L.; Shen, C.; and
  XingYu. 2026.
\newblock DeepEyes: Incentivizing ''Thinking with Images'' via Reinforcement
  Learning.
\newblock In \emph{The Fourteenth International Conference on Learning
  Representations}.

\bibitem[{Zhu et~al.(2025)Zhu, Wang, Chen, Liu, Ye, Gu, Tian, Duan, Su, Shao
  et~al.}]{zhu2025internvl3}
Zhu, J.; Wang, W.; Chen, Z.; Liu, Z.; Ye, S.; Gu, L.; Tian, H.; Duan, Y.; Su,
  W.; Shao, J.; et~al. 2025.
\newblock Internvl3: Exploring advanced training and test-time recipes for
  open-source multimodal models.
\newblock \emph{arXiv preprint arXiv:2504.10479}.

\end{thebibliography}

\clearpage
\onecolumn

\begin{center}
  {\LARGE\bfseries Technical Appendix of\\[3pt]
   TACO: Tool-Augmented Credit Optimization for Agentic Tool Use\par}
\end{center}
\vskip 1.0em

\noindent
This supplementary material provides additional details on \TACO{}, covering the training
algorithm, a theoretical motivation, the full experimental setup
(benchmarks, baselines, and data-curation pipelines), supplementary results, qualitative
case studies, and a probe-hacking analysis. All notation follows the main paper. The
appendix is organized as follows:

\begin{itemize}\setlength{\itemsep}{1pt}
  \item[] \textbf{A. TL;DR: Main Contributions and Takeaways}
  \item[] \textbf{B. Training Algorithm and Edge Cases}
  \begin{itemize}\setlength{\itemsep}{0pt}
    \item[] B.1. Training Algorithm
    \item[] B.2. Edge Cases
  \end{itemize}
  \item[] \textbf{C. Theoretical Motivation}
  \begin{itemize}\setlength{\itemsep}{0pt}
    \item[] C.1. Background: GRPO and Uniform Credit Assignment
    \item[] C.2. \DAPR{} as a Counterfactual Baseline
    \item[] C.3. Potential-Outcomes Interpretation of $\Dtool$
    \item[] C.4. Robustness to Probe-Hacking, Formally
    \item[] C.5. A Potential-Based-Shaping View of the Process Channel
    \item[] C.6. \OGAR{} as Conservative Advantage Masking
    \item[] C.7. The Two-Channel Objective
    \item[] C.8. Scope and Assumptions
  \end{itemize}
  \item[] \textbf{D. Experimental Setup Details}
  \begin{itemize}\setlength{\itemsep}{0pt}
    \item[] D.1. Benchmarks
    \item[] D.2. Baselines
    \item[] D.3. Implementation Details
    \item[] D.4. SFT Data Curation
    \item[] D.5. RL Data Curation
    \item[] D.6. Channel-Weight Sensitivity
  \end{itemize}
  \item[] \textbf{E. Show Cases}
  \item[] \textbf{F. Limitations and Future Work}
\end{itemize}

\appendix

\section{TL;DR: Main Contributions and Takeaways}
\noindent\textbf{Contributions.}
\begin{itemize}\setlength{\itemsep}{2pt}
  \item \textbf{\TACO{}}, a \textbf{GRPO} variant for code-tool visual agents that turns a
  single per-call signal into both a reward and a routing rule, coupling \DAPR{} and
  \OGAR{} into one objective.
  \item \textbf{\DAPR{}}, a self-supervised, judge-free per-tool-call reward: it scores a
  call by the before/after difference of the agent's own answer-probe outcomes, reusing
  the existing answer checker with no auxiliary model and at near-zero added cost.
  \item \textbf{\OGAR{}}, a parameter-free, token-level rule that routes the final-answer
  advantage only to the segments responsible for the outcome, suppressing wasted calls
  without any explicit cost term.
\end{itemize}

\medskip
\noindent\textbf{Takeaways.}
\begin{enumerate}\setlength{\itemsep}{2pt}
  \item \textbf{Credit the call, not the trajectory.} Tying each tool call's reward to its
  own measurable effect on the answer---rather than to the whole trajectory's
  outcome---is what lets the policy learn \emph{when} a crop helps and abstain when it
  does not.
  \item \textbf{Differencing beats absolute probes.} Because \DAPR{} subtracts the
  pre-tool baseline, it cancels what the model ``already knew'' and is naturally robust to
  probe-hacking, where an absolute probe score can be inflated by writing the answer early.
  \item \textbf{Routing matters as much as the reward.} A scalar tool value is not enough;
  \OGAR{} must deliver the final-answer advantage to the responsible tokens, otherwise
  redundant calls are over-credited and correct pre-tool reasoning is wrongly blamed.
  \item \textbf{Accuracy and efficiency together.} \TACO{} attains the best average among
  open-source models while invoking tools only when they help, so it is simultaneously the
  most accurate and the lowest-latency code-tool agent.
  \item \textbf{Backbone-agnostic.} The gains transfer from Qwen2.5-VL-7B to the stronger
  Qwen3-VL-8B, indicating they come from the credit-assignment mechanism rather than from
  the base model or data.
\end{enumerate}

\section{Training Algorithm and Edge Cases}
This section provides the implementation-level details omitted from the main paper:
pseudocode for the training loop and the edge cases the gate must handle. The notation
($\mathcal{T}_1,\mathcal{C},\mathcal{T}_2$, $\aone,\atwo,\Dtool,\Aone,\Atwo,m,g$) is that
of the main paper.

\subsection{Training Algorithm}
\TACO{} is the two-stage pipeline of Algorithm~\ref{alg:taco}: an SFT cold-start that
establishes the Think--Code--Answer format and fixes the reference policy
$\pi_{\mathrm{ref}}$, followed by group-relative RL with the gated dual-channel advantage.
Per prompt, each rollout is parsed into its three segments, the tool code is executed, the
two probes are decoded to score the call, and the accuracy and tool-value advantages are
group-normalized; the per-token gate $m$ and the trajectory gate $g$ then route the two
channels before the clipped-GRPO update.

\begin{algorithm}[t]
\caption{\TACO{}: two-stage training.}
\label{alg:taco}
\begin{algorithmic}[1]
\REQUIRE base VLM $\pi_\theta$; SFT set $\mathcal{D}_{\mathrm{sft}}$, RL prompts $\mathcal{D}_{\mathrm{rl}}$; checker $\ORM$; group size $G$; weights $\alpha_1,\alpha_2$; clip $\epsilon$, lr $\eta$
\STATE \textbf{Stage 1 (SFT).} fine-tune $\pi_\theta$ on $\mathcal{D}_{\mathrm{sft}}$; set $\pi_{\mathrm{ref}}\!\leftarrow\!\pi_\theta$, $\pi_{\theta_{\mathrm{old}}}\!\leftarrow\!\pi_\theta$
\STATE \textbf{Stage 2 (RL).}
\FOR{each mini-batch of prompts $(q,I)\in\mathcal{D}_{\mathrm{rl}}$}
  \STATE sample a group of $G$ rollouts $\{\tau^{(i)}\}_{i=1}^{G}\sim\pi_{\theta_{\mathrm{old}}}(\cdot\mid q,I)$
  \FOR{$i=1,\dots,G$}
    \STATE parse $\tau^{(i)}\!=\!(\mathcal{T}_1,\mathcal{C},\mathcal{T}_2,\Afin)$; execute $\mathcal{C}$ in the sandbox $\rightarrow$ observation $\imgk$
    \STATE \emph{pre-tool probe:} prefill \texttt{</think><answer>} after $\mathcal{T}_1$, greedy-decode $\aone$ from $(q,I,\mathcal{T}_1)$
    \STATE \emph{post-tool probe:} greedy-decode $\atwo$ from $(q,I,\mathcal{T}_1,\mathcal{C},\imgk,\mathcal{T}_2)$
    \STATE $\Dtool^{(i)}\!\leftarrow\!\ORM(\atwo)\!-\!\ORM(\aone)$;\quad $\Racc^{(i)}\!\leftarrow\!\ORM(\Afin)\!+\!0.5\,\Rfmt^{(i)}$
  \ENDFOR
  \STATE $\Aone^{(i)}\!\leftarrow\!(\Racc^{(i)}\!-\!\mu_{R})/\sigma_{R}$,\;\; $\Atwo^{(i)}\!\leftarrow\!(\Dtool^{(i)}\!-\!\mu_{\Delta})/\sigma_{\Delta}$ \hfill// group mean/std over $i{=}1{:}G$
  \FOR{$i=1,\dots,G$, token $t\in\tau^{(i)}$}
    \STATE set $m^{(i)}[t]\!=\!1$ if $t\!\in\!\mathcal{T}_2$, or $t\!\in\!\mathcal{C}$ with $\Dtool^{(i)}\!\neq\!0$, or $t\!\in\!\mathcal{T}_1$ with $\Dtool^{(i)}\!\geq\!0$; else $m^{(i)}[t]\!=\!0$
  \ENDFOR
  \STATE set $g^{(i)}\!=\!1$ if $\Dtool^{(i)}\!\geq\!0$, else $g^{(i)}\!=\!0$ \hfill// process channel off on misleading calls
  \STATE $\mathcal{L}\!\leftarrow\!\alpha_1\mathcal{L}_{\mathrm{GRPO}}(m\!\odot\!\Aone)+\alpha_2\mathcal{L}_{\mathrm{GRPO}}(g\,\Atwo)$ \hfill// clipped surrogate, Eq.~\eqref{eq:grpo-loss}
  \STATE $\theta\!\leftarrow\!\theta-\eta\nabla_\theta\mathcal{L}$;\quad periodically $\pi_{\theta_{\mathrm{old}}}\!\leftarrow\!\pi_\theta$ \hfill// KL weight $\beta=0$
\ENDFOR
\ENSURE trained policy $\pi_\theta$
\end{algorithmic}
\end{algorithm}

\subsection{Edge Cases}
\paragraph{No tool call.} If a rollout emits no \texttt{<code>} block and answers directly,
the tool branch is empty: there is no $\Dtool$ and the process channel is inactive, so only
the (gated) accuracy channel applies, with $\Aone$ on the answer tokens. This lets the
policy abstain from tools on items it already solves, at no penalty.

\paragraph{Multiple tool calls.} When a trajectory contains more than one call, the
post-tool probe is read after the \emph{final} call, so $\Dtool$ measures the value of the
entire tool branch (all calls, their observations, and the reasoning they trigger) against
the same pre-tool baseline $\aone$, and $\mathcal{C}$ is taken as the union of the code
blocks. Our experiments operate almost entirely in the single-call regime, for which the
before/after split is cleanest; finer per-call attribution within a multi-call branch is
left to future work.

\paragraph{Probe parse failure.} A probe whose decoded string cannot be parsed into a
valid answer is scored $\ORM=0$, so it neither rewards nor penalizes the call; this keeps
malformed probe decodes from injecting a spurious $\Dtool$.

\section{Theoretical Motivation}

This section motivates \TACO{} from first principles. We start from the GRPO objective
and its credit-assignment limitation (\ref{app:grpo}), recast \DAPR{} as a counterfactual
difference (\ref{app:dapr-theory}), give a potential-based-shaping reading of the process
channel (\ref{app:shaping}), show that \OGAR{} is a conservative masking of the outcome
advantage (\ref{app:ogar-theory}), and finally state precisely what is and is not claimed
(\ref{app:scope}).

\subsection{Background: GRPO and Uniform Credit Assignment}
\label{app:grpo}
For a prompt $x=(q,I)$, GRPO~\citep{deepseekmath} samples a group of $G$ trajectories
$\{\tau^{(i)}\}_{i=1}^{G}\sim\pi_{\theta_{\mathrm{old}}}$, scores each with a scalar reward
$R^{(i)}$, and forms the \emph{group-relative} advantage
\begin{equation}
A^{(i)} = \frac{R^{(i)} - \mu_R}{\sigma_R}, \quad
\mu_R = \tfrac{1}{G}\!\sum_{j} R^{(j)}, \;\;
\sigma_R = \mathrm{std}\big(\{R^{(j)}\}\big).
\label{eq:grpo-adv}
\end{equation}
Writing $\rho^{(i)}_t=\pi_\theta(o^{(i)}_t\mid x,o^{(i)}_{<t})/\pi_{\theta_{\mathrm{old}}}(\cdot)$
for the per-token importance ratio, the clipped surrogate is
\begin{equation}
\mathcal{L}_{\mathrm{GRPO}}(A)=\mathbb{E}_{i,t}\Big[\min\!\big(\rho^{(i)}_t A^{(i)},\,
\mathrm{clip}(\rho^{(i)}_t,1{-}\epsilon,1{+}\epsilon)\,A^{(i)}\big)\Big].
\label{eq:grpo-loss}
\end{equation}
The defining property of \eqref{eq:grpo-loss} is that $A^{(i)}$ is \emph{constant across
all tokens $t$} of $\tau^{(i)}$: the outcome reward is broadcast uniformly. Decompose a
code-tool trajectory into three segments, the pre-tool reasoning $\mathcal{T}_1$, the code
$\mathcal{C}$, and the post-tool reasoning with the final answer $\mathcal{T}_2$. Uniform
broadcasting gives the code tokens the gradient
$A^{(i)}\sum_{t\in\mathcal{C}}\nabla_\theta\log\pi_\theta(o^{(i)}_t)$, so $\mathcal{C}$ is
reinforced whenever the \emph{trajectory} ends correctly, regardless of whether the tool
call contributed; symmetrically, a correct $\mathcal{T}_1$ is penalized whenever a later
tool call spoils the answer. Outcome-level credit thus cannot separate a call's
contribution from the trajectory's overall correctness. \TACO{} addresses this with a
per-call reward (\DAPR{}) and a per-segment gate (\OGAR{}).

\subsection{\DAPR{} as a Counterfactual Baseline}
\label{app:dapr-theory}
Let $c_1=(q,I,\mathcal{T}_1)$ be the context just before the call and
$c_2=(q,I,\mathcal{T}_1,\mathcal{C},\imgk,\mathcal{T}_2)$ the context just after it. Each
probe prefills the answer header and greedily decodes
$a=\arg\max_a \pi_\theta(a\mid c)$, scored by the verifiable checker
$\ORM(\cdot)\in\{-1,0,+1\}$. The tool value is the difference
\begin{equation}
\Dtool = \ORM(\atwo) - \ORM(\aone).
\label{eq:dapr-theory}
\end{equation}
Equation~\eqref{eq:dapr-theory} is a \emph{counterfactual} estimate of the value of the
tool branch: $\ORM(\atwo)$ is the outcome with the branch and $\ORM(\aone)$ is the outcome
on the same trajectory \emph{without} it. Equivalently, $\ORM(\aone)$ serves as a
trajectory-specific reference for the post-tool outcome. Because $c_1$ is a prefix
of $c_2$, the subtraction removes the component of the outcome explained by the shared
pre-tool context (in particular any answer the model has already committed to in
$\mathcal{T}_1$), and isolates the \emph{marginal} effect of extending the context with
the tool branch $(\mathcal{C},\imgk,\mathcal{T}_2)$. This is exactly the term a uniform
outcome reward conflates with the call: $\Dtool>0$ only when the branch turns a wrong
pre-tool answer right, $\Dtool<0$ when it spoils a right one, and $\Dtool=0$ when it does
not move the outcome.

We deliberately do \emph{not} claim that $\Dtool$ isolates the visual observation $\imgk$
alone. The marginal increment bundles $\imgk$ with the extra reasoning $\mathcal{T}_2$
that the observation triggers, and $\mathcal{T}_2$ exists only on the post-tool side, so it
does not cancel. $\Dtool$ therefore estimates the value of \emph{taking the tool branch},
which is the quantity the policy actually controls when it decides to call a tool.

\subsection{Potential-Outcomes Interpretation of $\Dtool$}
\label{app:po}
It is useful to read \eqref{eq:dapr-theory} through the lens of potential outcomes. Treat
invoking the tool branch as a binary treatment $T\in\{0,1\}$ applied to a fixed pre-tool
state $c_1$, and define the two potential outcomes
\begin{equation}
Y(1)=\ORM(\text{answer}\mid \text{branch taken}),\quad
Y(0)=\ORM(\text{answer}\mid \text{branch withheld}).
\end{equation}
The post-tool probe realizes $Y(1)$ (it answers from $c_2$) and the pre-tool probe realizes
$Y(0)$ (it answers from $c_1$), \emph{on the same trajectory}, so $\Dtool=Y(1)-Y(0)$ reads
off both potential outcomes directly rather than estimating a missing counterfactual.
Averaged over the group and the data, $\Dtool$ is the average realized effect of the tool
branches the policy produces. Because both outcomes are read from the \emph{same} pre-tool
context $c_1$, the comparison is not confounded by the pre-tool state---the obstacle for
outcome-only credit, where a tool's effect and the trajectory's prior correctness are
entangled. Consistent with \ref{app:dapr-theory}, the quantity is the effect of the whole
branch (including $\mathcal{T}_2$), not of the observation in isolation.

\subsection{Robustness to Probe-Hacking, Formally}
\label{app:hack-theory}
A generative probe is vulnerable to \emph{answer leakage}: the policy could write its
answer early into $\mathcal{T}_1$ so that the prefilled-header decode merely copies it. We
formalize why differencing neutralizes this.

\medskip
\noindent\textbf{Proposition~1 (Invariance to common-mode shifts).}
\emph{Suppose a change to $\mathcal{T}_1$ shifts both probe scores by a common amount,
$\ORM(\aone)\!\mapsto\!\ORM(\aone)+\delta$ and
$\ORM(\atwo)\!\mapsto\!\ORM(\atwo)+\delta$. Then $\Dtool$ is unchanged. In particular, if
the change makes both probes copy the same pre-committed answer $\hat y$
(so $\aone=\atwo=\hat y$), then $\Dtool=0$.}

\smallskip
\noindent\textit{Proof.} $\Dtool=\ORM(\atwo)-\ORM(\aone)$; adding $\delta$ to both terms
leaves the difference unchanged. If $\aone=\atwo=\hat y$ then
$\ORM(\atwo)=\ORM(\aone)=\ORM(\hat y)$, so $\Dtool=0$. \hfill$\square$

\smallskip
Answer leakage is exactly a common-mode shift of the pre-tool baseline (it raises
$\ORM(\aone)$ and $\ORM(\atwo)$ together), so by Proposition~1 the process channel grants it
no advantage: the only way to earn $\Dtool>0$ is to make the post-tool answer correct when
the pre-tool answer was not, which requires the tool branch to change the outcome. An
absolute-probe reward such as $\ORM(\aone)+\ORM(\atwo)$ lacks this invariance---under the
same shift it increases by $2\delta$---which is precisely the mechanism behind the
probe-hacking we observe for the additive-probe variant.

\subsection{A Potential-Based-Shaping View of the Process Channel}
\label{app:shaping}
Let $s_1$ and $s_2$ denote the reasoning states reached after $\mathcal{T}_1$ and after the
tool branch, and define the potential $\Phi(s)=\ORM(\mathrm{probe}(s))$ as the probe
correctness at a state. Then \eqref{eq:dapr-theory} is a potential difference,
\begin{equation}
\Dtool = \Phi(s_2) - \Phi(s_1).
\end{equation}
Potential-based reward shaping adds to the reward a term $F(s,s')=\gamma\,\Phi(s')-\Phi(s)$
and preserves the set of optimal policies~\citep{ng1999policy}; with $\gamma=1$ the process
signal $\Dtool$ has precisely this form.

\medskip
\noindent\textbf{Proposition~2 (Policy invariance of potential shaping).}
\emph{In an episodic MDP with reward $r$ and $\Phi(\text{terminal})=0$, replacing $r$ by
$r'(s,a,s')=r(s,a,s')+\gamma\Phi(s')-\Phi(s)$ leaves the optimal policy set unchanged, and
the optimal action values satisfy $Q'^{*}(s,a)=Q^{*}(s,a)-\Phi(s)$.}

\smallskip
\noindent\textit{Proof.} Along any trajectory the shaping terms telescope:
$\sum_{t\ge0}\gamma^{t}\big(\gamma\Phi(s_{t+1})-\Phi(s_t)\big)=-\Phi(s_0)$, using
$\Phi(\text{terminal})=0$. The shaped return thus differs from the original by the constant
$-\Phi(s_0)$, which does not depend on the policy; hence $\arg\max$ over policies is
preserved and $Q'^{*}(s,a)=Q^{*}(s,a)-\Phi(s)$. \hfill$\square$

\smallskip
The process channel is the $\gamma=1$ instance with $\Phi(s)=\ORM(\mathrm{probe}(s))$.
Proposition~2 is what licenses adding $\Dtool$ as an auxiliary signal: it changes
\emph{how fast} and \emph{where} credit flows, not the optimum implied by the outcome
reward. We use this for the shaping \emph{form}: in practice the channel is applied at the
group/trajectory level under the GRPO normalization of \eqref{eq:grpo-adv} rather than as an
exact per-step MDP potential, and the probe only approximates the state value, so we treat
it as a soundness argument rather than an end-to-end invariance guarantee.

\subsection{\OGAR{} as Conservative Advantage Masking}
\label{app:ogar-theory}
Standard GRPO sets the per-token advantage to $A[t]=\Aone$ for every $t$. \OGAR{} replaces
this with $A[t]=m[t]\,\Aone$, where $m[t]\in\{0,1\}$ is the outcome-conditioned gate of the
main paper. Since $m[t]$ only zeroes (never sign-flips) the advantage, the gated gradient
\begin{equation}
\nabla_\theta\mathcal{L} \;=\; \sum_{t} m[t]\,\Aone\,\nabla_\theta\log\pi_\theta(o_t)
\end{equation}
is a \emph{sub-sum} of the original GRPO gradient: credit is withheld, never inverted.

\medskip
\noindent\textbf{Proposition~3 (Conservativeness of gating).}
\emph{For $m[t]\in\{0,1\}$, the gated gradient is the sum of a subset of the GRPO
gradient's per-token terms. Gating never reverses the sign of any token's contribution; it
only removes it. Consequently \OGAR{} can withhold credit but cannot convert
outcome-justified reinforcement into penalization, or vice versa.}

\smallskip
\noindent\textit{Proof.} Multiplying a term by $m[t]\in\{0,1\}$ either keeps it ($m[t]=1$)
or sets it to $\mathbf 0$ ($m[t]=0$); neither operation flips its sign. The gated gradient
is $\sum_{t:\,m[t]=1}\Aone\,\nabla_\theta\log\pi_\theta(o_t)$, a subset of the ungated sum.
\hfill$\square$

\smallskip
The update therefore never pushes a token in the direction opposite to its
outcome-justified one; it only declines to assign credit on segments that the call's
outcome shows are not responsible. Two consequences follow directly from the gate
definition. On a \emph{right-but-redundant} call the code's advantage is set to $0$ rather
than positive, removing the gradient that would otherwise reward an unnecessary call and
thereby discouraging over-calling at no extra cost term. On a \emph{misleading} call the
blame is kept off the correct pre-tool reasoning $\mathcal{T}_1$, so a single spoiled tool
call does not teach the model to distrust reasoning that was already right.

\subsection{The Two-Channel Objective}
\label{app:twochannel}
\TACO{} combines the gated accuracy channel and the process channel into
\begin{equation}
\mathcal{L}_{\mathrm{TACO}}=\alpha_1\,\mathcal{L}_{\mathrm{GRPO}}(m\odot\Aone)
+\alpha_2\,\mathcal{L}_{\mathrm{GRPO}}(g\,\Atwo),
\label{eq:twochannel}
\end{equation}
where $\Aone$ is the group-normalized accuracy advantage \eqref{eq:grpo-adv} of
$\Racc=\ORM(\Afin)+0.5\,\Rfmt$, the mask $m$ restricts it to outcome-responsible tokens,
$\Atwo$ is the group-normalized advantage of $\Dtool$, and the trajectory gate $g$
(equal to $1$ when $\Dtool\ge0$ and $0$ otherwise) disables the process channel on
misleading calls. The two
channels act on \emph{different supports}: $m\odot\Aone$ lives on a subset of tokens
(\ref{app:ogar-theory}), while $g\,\Atwo$, when active, applies to the whole sequence. This
separation is deliberate. On a misleading call the penalty is already carried by the gated
accuracy channel on $\mathcal{C}$ and $\mathcal{T}_2$; setting $g=0$ there avoids
double-counting it through the process channel, and, since $m$ keeps $\Aone$ off
$\mathcal{T}_1$, the correct pre-tool reasoning is left unpenalized by \emph{both} channels.
The coefficients trade outcome correctness ($\alpha_1$) against tool-value shaping
($\alpha_2$); the strong asymmetry we use ($\alpha_1{=}1.0$, $\alpha_2{=}0.15$) keeps the
verifiable outcome as the primary objective and treats $\Dtool$ as an auxiliary shaping
signal (\ref{app:shaping}).

\subsection{Scope and Assumptions}
\label{app:scope}
The analysis above relies on three assumptions, which we state plainly.
(i)~\emph{Faithful probing}: the prefilled-header greedy decode reflects the answer the
model would commit to from the given context. (ii)~\emph{Verifiable outcomes}: $\ORM$ is a
rule-based checker, so $\Dtool\in\{-2,-1,0,1,2\}$ and the method applies most directly to
tasks with checkable answers. (iii)~\emph{Single decisive call}: the before/after split is
cleanest when a trajectory contains one tool branch (the regime our setting targets).
Finally, we make \emph{no} variance-reduction claim: differencing two probe scores can
\emph{increase} per-sample variance relative to a single outcome reward, and we rely on the
group normalization in \eqref{eq:grpo-adv} to control it; the empirical training-reward
dynamics, not a variance inequality, are our evidence that the signal is well-behaved.

\section{Experimental Setup Details}

\subsection{Benchmarks}
We evaluate on twelve benchmarks spanning three groups (Table~\ref{tab:benchmarks}). For
each we report accuracy ($\%$) on the standard evaluation split, and the macro-average is
taken over all twelve. Answers are extracted and scored with the default
\citet{vlmevalkit} protocol, identical for every model.

\begin{table}[t]
\centering
\caption{Evaluation benchmarks grouped by category, with the number of questions in each
benchmark.}
\label{tab:benchmarks}
\small
\setlength{\tabcolsep}{6pt}
\renewcommand{\arraystretch}{1.15}
\begin{tabular}{@{}l l r@{}}
\toprule
\textbf{Benchmark} & \textbf{Category} & \textbf{\#Samples} \\
\midrule
V$^{*}$~\citep{vstar}            & \multirow{4}{*}{Perception} & $191$ \\
HR-Bench-4K~\citep{hr4k}         &                             & $800$ \\
HR-Bench-8K~\citep{hr4k}         &                             & $800$ \\
MME-RealWorld~\citep{zhang2025mme} &                           & $29{,}429$ \\
\midrule
MathVision~\citep{MathVision}    & \multirow{5}{*}{Reasoning}  & $3{,}040$ \\
MathVista~\citep{lu2024mathvista}&                             & $6{,}141$ \\
MathVerse~\citep{zhang2024mathverse} &                         & $2{,}612$ \\
WeMath~\citep{qiao2025we}        &                             & $6{,}500$ \\
LogicVista~\citep{xiao2024logicvista} &                        & $448$ \\
\midrule
MMStar~\citep{chen2024we}        & \multirow{3}{*}{General}    & $1{,}500$ \\
ChartQA~\citep{masry2022chartqa} &                             & $2{,}500$ \\
BLINK~\citep{fu2024blink}        &                             & $3{,}807$ \\
\bottomrule
\end{tabular}
\end{table}

\paragraph{Perception.}
These benchmarks stress fine-grained recognition, often on high-resolution images where a
decisive detail is small and easy to miss --- exactly the regime a crop/zoom tool targets.
\begin{itemize}\setlength{\itemsep}{1pt}
  \item \textbf{HR-Bench-4K / HR-Bench-8K}~\citep{hr4k} evaluate perception on
  high-resolution images at $4$K and $8$K, each with $800$ samples covering
  fine-grained single-instance and cross-instance perception (attributes, positions,
  and relations of small objects). We report the two resolutions as separate columns.
  \item \textbf{MME-RealWorld}~\citep{zhang2025mme} is a large-scale, manually annotated
  real-world benchmark of high-resolution images with deliberately challenging
  perception and reasoning questions.
  \item \textbf{V$^{*}$}~\citep{vstar} is a visual-search benchmark of $191$ items that
  requires locating a small target in a high-resolution scene, with attribute-recognition
  and spatial-relationship subtasks.
\end{itemize}

\paragraph{Reasoning.}
These benchmarks test multimodal mathematical and logical reasoning over diagrams,
figures, and charts.
\begin{itemize}\setlength{\itemsep}{1pt}
  \item \textbf{MathVision}~\citep{MathVision} contains $3{,}040$ competition-level visual
  math problems across $16$ disciplines and five difficulty levels.
  \item \textbf{MathVista}~\citep{lu2024mathvista} aggregates $6{,}141$ examples of
  mathematical reasoning in visual contexts from $28$ source datasets plus three newly
  collected ones (IQTest, FunctionQA, PaperQA).
  \item \textbf{MathVerse}~\citep{zhang2024mathverse} provides $2{,}612$ diagram-based math
  problems, each rendered in several versions that shift information between the text and
  the diagram to test genuine visual understanding.
  \item \textbf{WeMath}~\citep{qiao2025we} organizes visual math problems into a hierarchy
  of $67$ knowledge concepts, probing reasoning beyond end-to-end accuracy.
  \item \textbf{LogicVista}~\citep{xiao2024logicvista} comprises $448$ multiple-choice
  questions evaluating logical reasoning (inductive, deductive, spatial, and more) in
  visual contexts.
\end{itemize}

\paragraph{General.}
These benchmarks measure broad multimodal understanding.
\begin{itemize}\setlength{\itemsep}{1pt}
  \item \textbf{MMStar}~\citep{chen2024we} is a $1{,}500$-sample, human-curated benchmark
  of vision-indispensable questions spanning six core capabilities and $18$ axes, built to
  reduce text-only solvability and data leakage.
  \item \textbf{ChartQA}~\citep{masry2022chartqa} tests question answering over charts that
  requires both visual reading and logical/arithmetic reasoning, combining human-written
  and machine-generated questions.
  \item \textbf{BLINK}~\citep{fu2024blink} contains $3{,}807$ multiple-choice questions
  over $14$ classic visual-perception tasks that are easy for humans but remain hard for
  current MLLMs.
\end{itemize}

\subsection{Baselines}
We compare against three families. \emph{Closed-source} models: GPT-4o and
Gemini-2.5-Pro. \emph{Open-source MLLMs without visual tools}: Qwen2.5-VL,
Qwen2.5-VL-32B~\citep{bai2025qwen25vltechnicalreport},
InternVL3~\citep{zhu2025internvl3}, LLaVA-OneVision~\citep{li2024llava}, and
Qwen3-VL~\citep{bai2025qwen3}. The most relevant family is the \emph{7--8B code-tool /
visual-agent models}, which, like \TACO{}, act on the image through code or visual
operations and then reason over the result:

\begin{itemize}\setlength{\itemsep}{2pt}
  \item \textbf{Thyme}~\citep{thyme} (\emph{Think Beyond Images}) goes beyond simple
  cropping to a broad space of image-processing operations expressed as code (zoom,
  rotation, contrast, and general computation), trained with a two-stage SFT-then-RL
  recipe. It is the source of our SFT corpus and our base recipe.
  \item \textbf{DeepEyes}~\citep{deepeyes} incentivizes ``thinking with images'' through
  end-to-end reinforcement learning, where a zoom-in tool-use behavior emerges natively
  without pre-collected reasoning data or an external tool model.
  \item \textbf{DeepEyesV2}~\citep{deepeyesv2} builds an agentic multimodal model with a
  cold-start stage followed by RL, observing that RL alone fails to induce robust tool
  use; it exhibits task-adaptive invocation, using image operations for perception and
  numerical computation for reasoning.
  \item \textbf{Pixel-Reasoner}~\citep{pixelreasoner} equips a VLM with pixel-space
  operations (zoom-in, select-frame) and uses curiosity-driven RL to escape the
  ``learning trap'' where the model falls back on text-only reasoning and neglects the
  visual operations.
  \item \textbf{Mini-o3}~\citep{minio3} scales up multi-turn visual search to tens of
  interaction turns, using a visual-probe dataset and an over-turn masking strategy that
  avoids penalizing long trajectories during RL.
  \item \textbf{MathCoder-VL}~\citep{mathcodervl} bridges vision and code for multimodal
  mathematical reasoning, aligning visual content with executable code to improve
  diagram-grounded math problem solving.
  \item \textbf{CodeV}~\citep{codev} represents visual tools as executable Python code and
  trains with Tool-Aware Policy Optimization, a process-level RL framework that assigns
  dense rewards directly on tool inputs and outputs (via an external judge) to encourage
  faithful, evidence-consistent tool use.
  \item \textbf{PyVision}~\citep{pyvision} is an agentic framework in which the MLLM
  autonomously generates, executes, and refines task-specific Python tools at inference,
  enabling flexible and interpretable visual problem solving.
\end{itemize}

\subsection{Implementation Details}
Following Thyme, we build on Qwen2.5-VL-7B with $2$ epochs of SFT followed by $1$ epoch of
GRPO. We set $\alpha_1 = 1.0$ and $\alpha_2 = 0.15$, use no KL penalty ($\beta = 0$;
i.e.\ KL regularization is disabled), and sample $G = 8$ rollouts per prompt at
temperature $1.0$, with a total batch size of $128$ and learning rate $1\times10^{-6}$.
Training runs on a single node of $8\times$80\,GB A100 GPUs. We report the sensitivity to
the channel weights $\alpha_1,\alpha_2$ in Table~\ref{tab:alpha}.

\paragraph{Prompt templates.}
Both stages use the same system prompt (Table~\ref{tab:sysprompt}), inherited from
Thyme~\citep{thyme}: it instructs the agent to reason step by step and, optionally, to emit
sandboxed Python for image manipulation, returning the processed image or result for
further reasoning. The per-example user prompt (Table~\ref{tab:userprompt}) supplies the
image together with the question and the image path and size, and fixes the required
\texttt{<think>}/\texttt{<answer>} output format. Both are applied to every $(q,I)$ under
the Qwen2.5-VL chat template.

\begin{table}[t]
\centering
\caption{System prompt used for the SFT cold-start (and reused unchanged in RL). Tag and
code tokens are shown in \texttt{typewriter}.}
\label{tab:sysprompt}
\small
\renewcommand{\arraystretch}{1.25}
\begin{tabular}{@{}p{0.96\columnwidth}@{}}
\toprule
\textbf{System Prompt (SFT \& RL)} \\
\midrule
You are a helpful assistant. \\[4pt]
Solve the following problem step by step, and optionally write Python code for image
manipulation to enhance your reasoning process. The Python code will be executed by an
external sandbox, and the processed image or result (wrapped in
\texttt{\textless sandbox\_output\textgreater\textless/sandbox\_output\textgreater}) can be
returned to aid your reasoning and help you arrive at the final answer. \\[4pt]
\textbf{Reasoning \& Image Manipulation (Optional but Encouraged):} \\[2pt]
$\bullet$~You have the capability to write executable Python code to perform image
manipulations (e.g., cropping to a Region of Interest (ROI), resizing, rotation, adjusting
contrast) or perform calculation for better reasoning. \\[2pt]
$\bullet$~The code will be executed in a secure sandbox, and its output will be provided
back to you for further analysis. \\[2pt]
$\bullet$~All Python code snippets \textbf{must} be wrapped as follows: \\[2pt]
\hspace*{1em}\texttt{\textless code\textgreater} \\
\hspace*{1em}\texttt{\textasciigrave\textasciigrave\textasciigrave python} \\
\hspace*{1em}\texttt{\#\ your code.} \\
\hspace*{1em}\texttt{\textasciigrave\textasciigrave\textasciigrave} \\
\hspace*{1em}\texttt{\textless/code\textgreater} \\[2pt]
$\bullet$~At the end of the code, print the path of the processed image
(\texttt{processed\_path}) or the result for further processing in a sandbox environment. \\
\bottomrule
\end{tabular}
\end{table}

\begin{table}[t]
\centering
\caption{Per-example user prompt used for the SFT cold-start (and reused unchanged in RL).
\texttt{[$\cdot$]} placeholders are filled per sample; \texttt{<image>} is the visual token
consumed by the Qwen2.5-VL chat template. Tag tokens are shown in \texttt{typewriter}.}
\label{tab:userprompt}
\small
\renewcommand{\arraystretch}{1.25}
\begin{tabular}{@{}p{0.96\columnwidth}@{}}
\toprule
\textbf{User Prompt (SFT \& RL)} \\
\midrule
\texttt{<image>} \\[4pt]
User's Question: \texttt{[User Question]} \\
User Image Path: \texttt{[Image Path]} \\
User Image Size: \texttt{[Image Size]} \\[4pt]
Output Format (strict adherence required): \\[2pt]
\hspace*{1em}\texttt{<think>}Your detailed reasoning process, including any code, should go here.\texttt{</think>} \\
\hspace*{1em}\texttt{<answer>}Your final answer to the user's question goes here.\texttt{</answer>} \\
\bottomrule
\end{tabular}
\end{table}

\subsection{SFT Data Curation}
\label{app:sft}
Our SFT data is built on the Thyme SFT corpus~\citep{thyme}, re-curated with three
filters. \textbf{(i) Execution validity}: we re-run every code block in our sandbox and
discard trajectories with execution errors or with tool observations/answers
inconsistent with the actual output, which would otherwise teach the model to
hallucinate observations. \textbf{(ii) Tool necessity}: we drop samples that
Qwen2.5-VL-7B~\citep{bai2025qwen25vltechnicalreport} already solves without tools
(pass@$8 = 1$), keeping only trajectories where a tool call is genuinely needed.
\textbf{(iii) Quality}: Gemini-3-Pro scores each trajectory for reasoning coherence and
tool-use rationale, and low-quality or blind-tool-use traces are removed.

\subsection{RL Data Curation}
For RL we follow CodeV~\citep{codev}, building on its open-source prompt data and
adopting its data-cleaning recipe. Keeping only questions with verifiable ground-truth
answers, we clean them in two ways. \textbf{(i) Environmental fidelity}: each prompt is
passed through Gemini-3-Pro to check image quality, question clarity, and image--text
consistency, and prompts with corrupted images or severe ambiguity are removed.
\textbf{(ii) Difficulty calibration}: prompts that our SFT checkpoint already solves on
all $G = 8$ rollouts yield zero-variance accuracy rewards (no GRPO advantage), so we
remove them.

\subsection{Channel-Weight Sensitivity}
We fix the accuracy weight $\alpha_1{=}1.0$ and sweep the tool-value weight $\alpha_2$,
reporting the macro-average over all twelve benchmarks (Table~\ref{tab:alpha}). The setting
$\alpha_1{=}1.0,\alpha_2{=}0.15$ is the configuration used throughout the paper. Small
$\alpha_2$ keeps the verifiable outcome as the dominant objective while still letting the
tool-value channel shape exploration; very large $\alpha_2$ would over-weight the auxiliary
signal relative to final correctness.

\begin{table}[t]
\centering
\caption{Sensitivity to the channel weights $\alpha_1,\alpha_2$: macro-average accuracy
over the twelve benchmarks. The row used in the paper is in \textbf{bold}.}
\label{tab:alpha}
\small
\setlength{\tabcolsep}{10pt}
\renewcommand{\arraystretch}{1.2}
\begin{tabular}{@{}c c c@{}}
\toprule
$\alpha_1$ & $\alpha_2$ & Avg.\ (12 benchmarks) \\
\midrule
$1.0$ & $0.05$ & 67.8 \\
$1.0$ & $0.10$ & 68.0 \\
\textbf{1.0} & \textbf{0.15} & \textbf{68.1} \\
$1.0$ & $0.30$ & 67.9 \\
$1.0$ & $0.50$ & 67.7 \\
\bottomrule
\end{tabular}
\end{table}

\section{Show Cases}
\begin{figure}[p]\centering
\includegraphics[width=\textwidth]{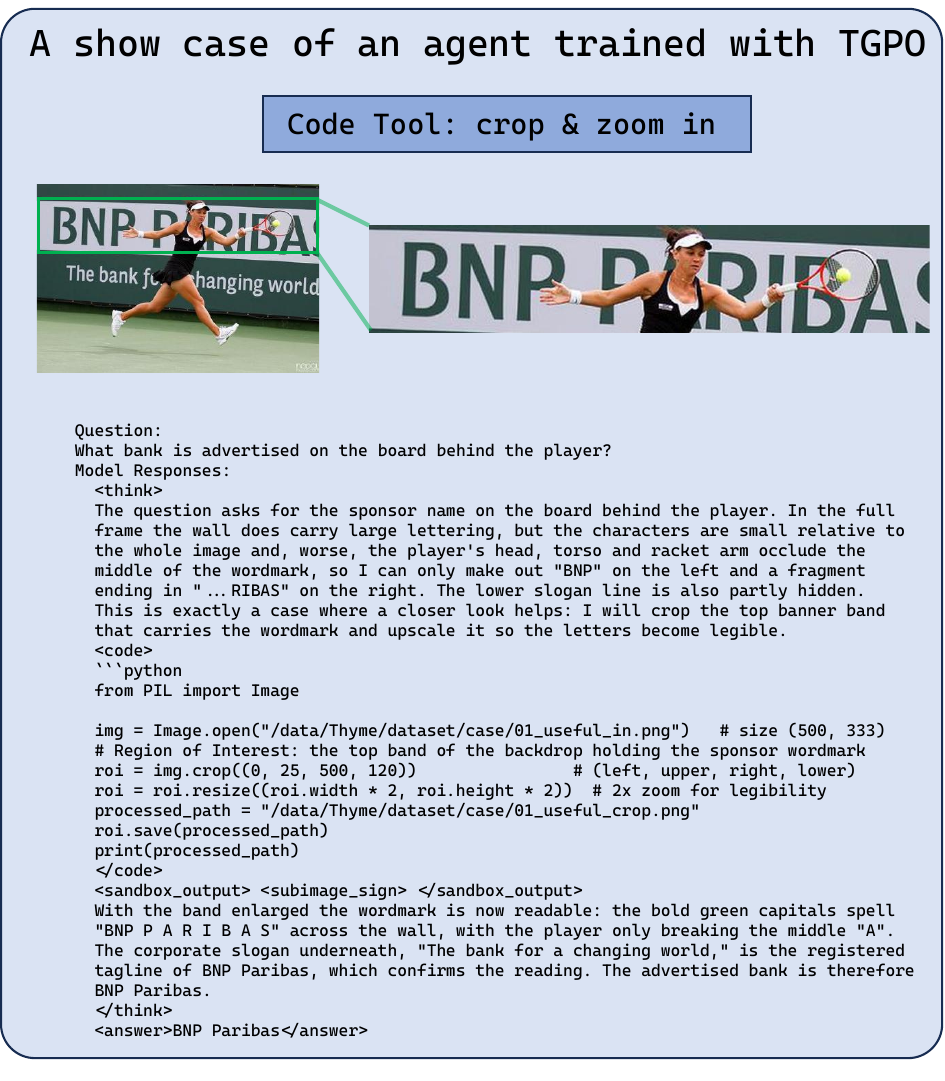}
\caption{Reading an occluded sponsor wordmark: the agent crops the backdrop banner and
zooms $2\times$ to recover ``BNP Paribas''.}
\label{fig:sc1}
\end{figure}

\begin{figure}[p]\centering
\includegraphics[width=\textwidth]{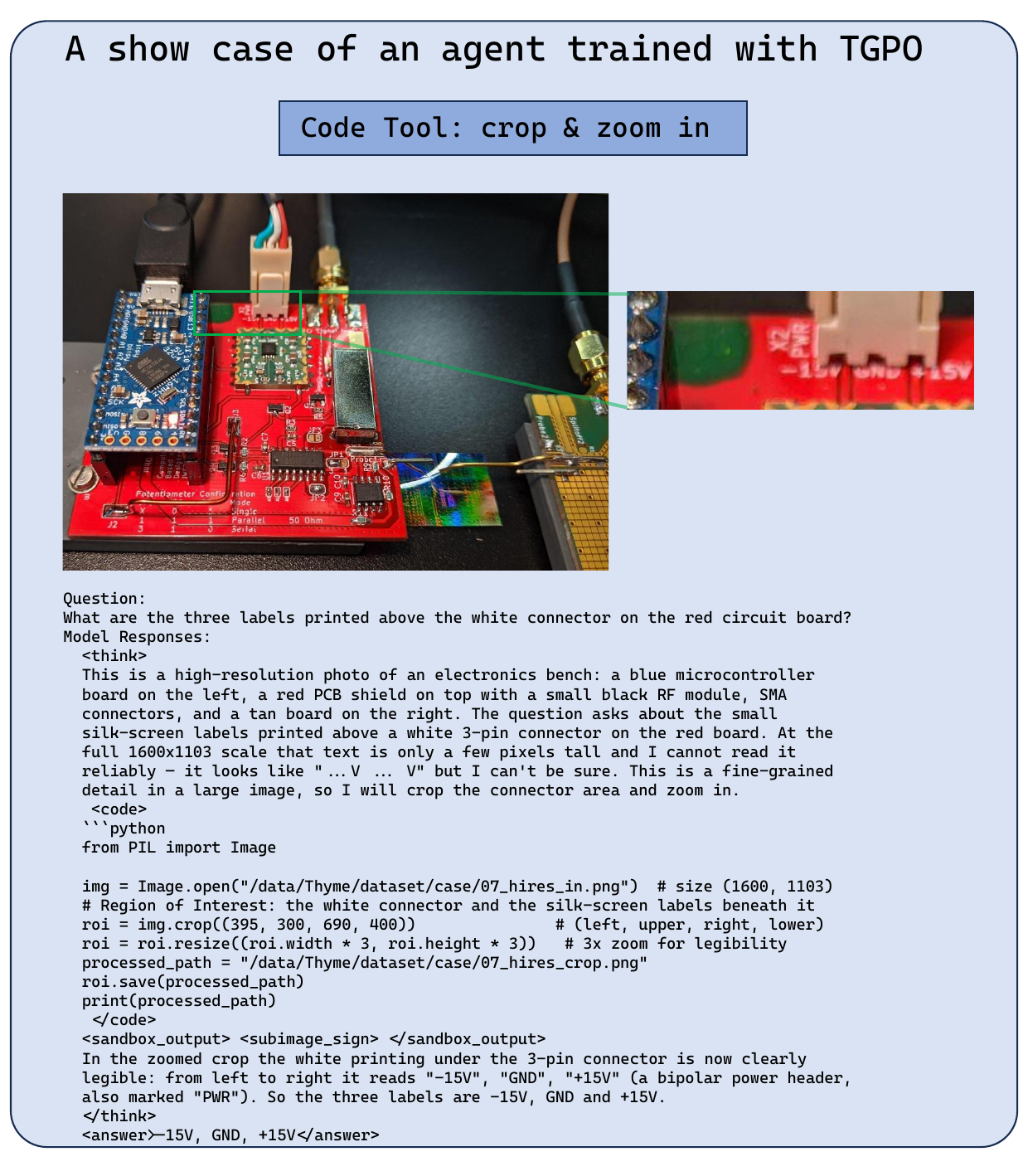}
\caption{High-resolution perception: the agent crops and zooms a circuit board to read the
three small labels printed above a connector.}
\label{fig:sc2}
\end{figure}

\begin{figure}[p]\centering
\includegraphics[width=\textwidth]{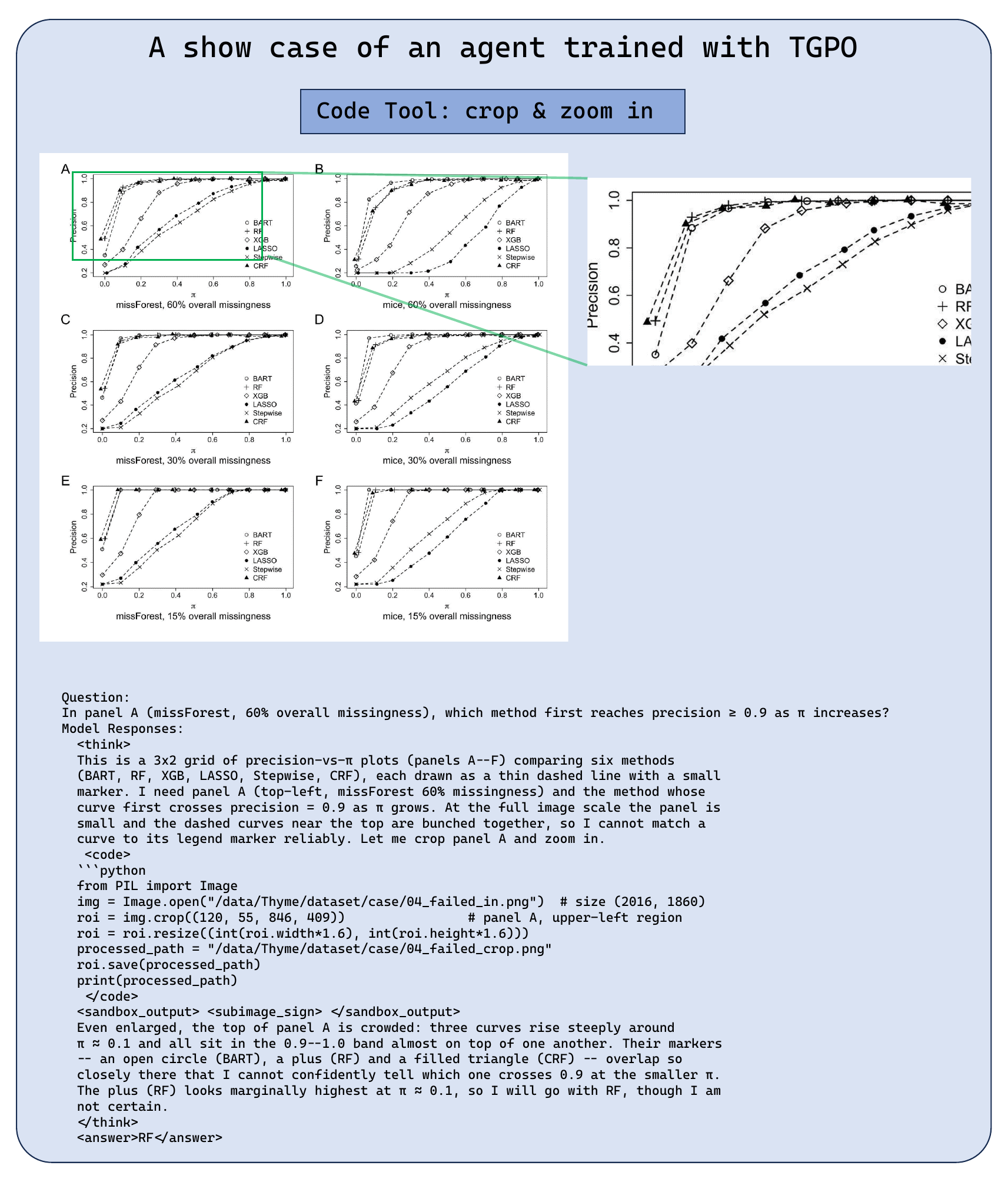}
\caption{Scientific-figure reading: the agent isolates one panel of a six-method
precision plot to extract the requested value.}
\label{fig:sc3}
\end{figure}

\begin{figure}[p]\centering
\includegraphics[width=\textwidth]{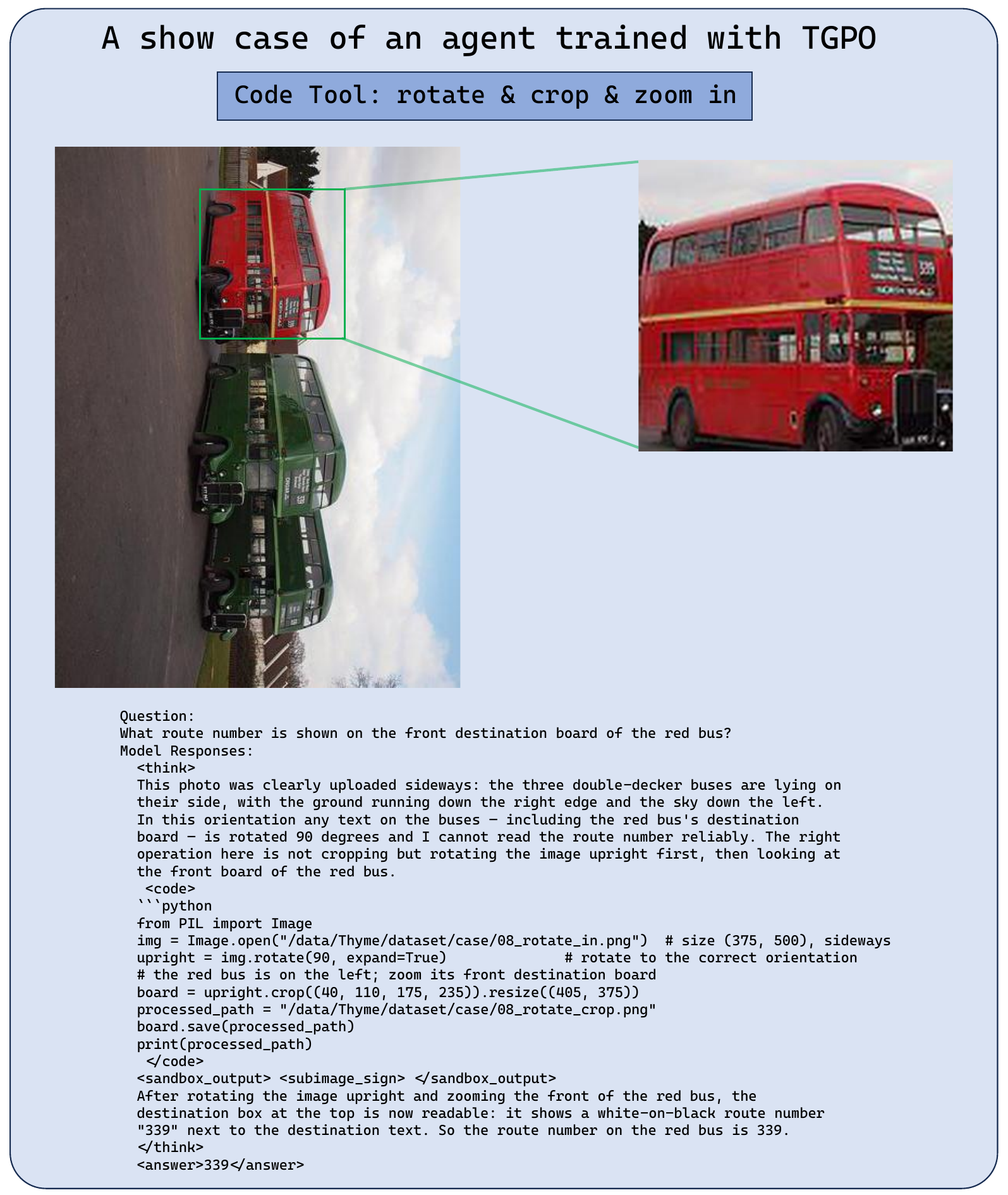}
\caption{Image rotation: the photo was uploaded sideways, so the agent rotates it before
reading the route number on the bus's destination board.}
\label{fig:sc4}
\end{figure}

\begin{figure}[p]\centering
\includegraphics[width=\textwidth]{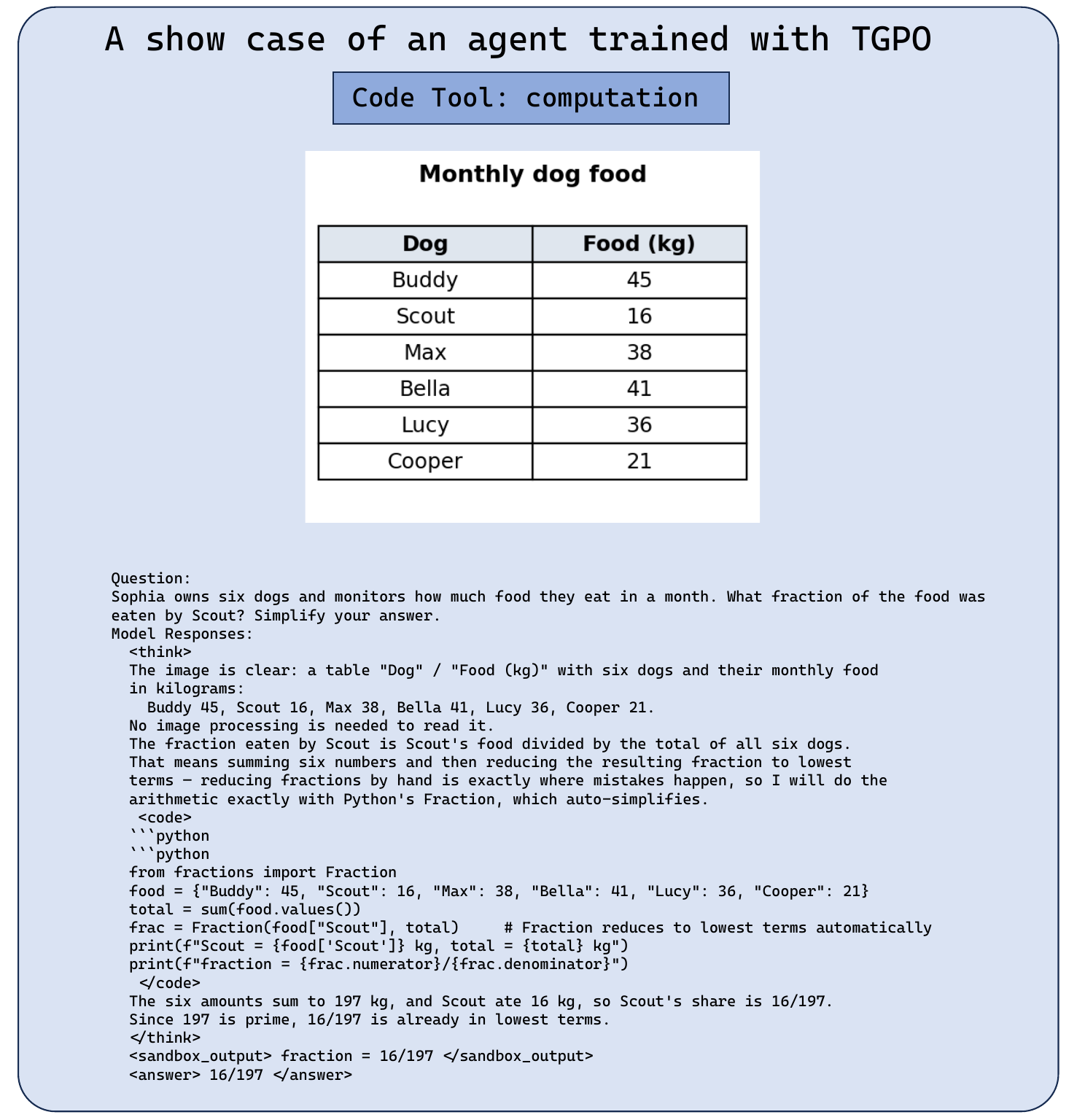}
\caption{Chart-grounded math: the agent crops the relevant chart region and computes the
requested fraction.}
\label{fig:sc5}
\end{figure}

Figures~\ref{fig:sc1}--\ref{fig:sc5} present five real trajectories from a \TACO{}-trained
agent, spanning crop/zoom perception, scientific-figure reading, image rotation, and
chart-grounded math. Each panel shows the question, the agent's reasoning, the Python tool
call and the view it returns, and the final answer.

\section{Limitations and Future Work}
\TACO{} relies on a rule-based outcome checker, so it applies most directly to tasks with
verifiable answers, and its probes assume the tool's effect is observable in the answer.
The single-call scoping exploits a clean before/after split; extending the
probe-difference signal to multi-call trajectories, open-ended generation, and richer
tool spaces is left to future work. Several further directions are promising: integrating
\TACO{} with model-compression methods for efficient deployment~\citep{feng2025dress,feng2026two},
coupling it with thought-augmented reasoning paradigms~\citep{wu2026beyond,wu2025thought,wu2026astar,wu2025pandora},
and pairing it with model-routing methods~\citep{jin2025radialrouter,jin2026exploring}.

\end{document}